\documentclass[pdflatex,sn-mathphys-num]{sn-jnl}

\usepackage{graphicx}%
\usepackage{multirow}%
\usepackage{amsmath,amssymb,amsfonts}%
\usepackage{amsthm}%
\usepackage{mathrsfs}%
\usepackage[title]{appendix}%
\usepackage{xcolor}%
\usepackage[table]{xcolor}
\usepackage{textcomp}%
\usepackage{manyfoot}%
\usepackage{booktabs}%
\usepackage{algorithm}%
\usepackage{algorithmicx}%
\usepackage{algpseudocode}%
\usepackage{listings}%
\usepackage{adjustbox}
\usepackage{makecell}
\usepackage{subcaption} 

\usepackage{comment}
\usepackage{url}
\usepackage{array}

\theoremstyle{thmstyleone}%
\theoremstyle{thmstyletwo}%

\theoremstyle{thmstylethree}%

\begin{document}


\title{From Interaction to Demonstration Quality in Virtual Reality: Effects of Interaction Modality and Visual Representation on Everyday Tasks}





\author*[1]{\fnm{Robin} \sur{Beierling}}\email{robin.helmert@uni-bielefeld.de}


\author[1]{\fnm{Manuel} \sur{Scheibl}}\email{manuel.scheibl@uni-bielefeld.de}
\author[2]{\fnm{Jonas} \sur{Dech}}\email{jdech@uni-bremen.de}
\author[]{\fnm{Abhijit} \sur{Vyas}}\email{vyasabhijit21@gmail.com}
\author[1]{\fnm{Anna-Lisa} \sur{Vollmer}}\email{anna-lisa.vollmer@uni-bielefeld.de}

\affil[1]{\orgdiv{Medical School OWL}, \orgname{Bielefeld University}, \city{Bielefeld}, \country{Germany}}
\affil[2]{\orgdiv{AICOR Institute for Artificial Intelligence}, \orgname{University of Bremen}, \city{Bremen}, \country{Germany}}


\abstract{Virtual Reality (VR) is increasingly used for training and demonstration purposes including a variety of applications ranging from robot learning to rehabilitation.
However, the choice of input device and its visualization might influence workload and thus user performance leading to suboptimal demonstrations or reduced training effects.
This study investigates how different VR input configurations—motion capture gloves, controllers with hand visualization, and controllers with controller visualization—affect user experience and task execution, with the goal of identifying which configuration is best suited for which type of task.
Participants performed various kitchen-related activities of daily living (ADLs), including object placement, cutting, cleaning, and pouring in a simulated environment.
To address two research questions, we evaluated user experience using the System Usability Scale and NASA Task Load Index (RQ1), and  task-specific interaction behavior (RQ2).
The latter was assessed using trajectory segmentation, analyzing movement efficiency, unnecessary actions, and execution precision.
While no significant differences in overall usability and workload were found, trajectory analysis revealed configuration-specific execution behaviors with different movement strategies.
Controllers enabled significantly faster task completion with less movement variability in pick-and-place style tasks such as table setting. In contrast, motion capture gloves produced more natural movements with fewer unnecessary actions, but also showed greater variance in movement patterns for manner-oriented tasks such as cutting bread.
These findings highlight trade-offs between efficiency and naturalism, and have implications for optimizing VR-based training, improving the quality of user-generated demonstrations, and tailoring interaction design to specific application goals.
}




\maketitle

\section{Introduction}\label{sec1}

\label{sec:introduction}

Virtual Reality (VR) has revolutionized the way we interact with digital environments, offering immersive and interactive spaces tailored to specific needs.
It enables the simulation of high-fidelity scenarios for various applications, including robot learning through trajectory recording~\cite{burghardt2020programming,DBLP:journals/presence/HamonLRR11,DBLP:conf/petra/TheofanidisSLM17}, training~\cite{aim2016effectiveness,DBLP:journals/cbsn/MantovaniCGR03,gurusamy2009virtual}, and rehabilitation~\cite{levin2015emergence,warnier2020effect,liu2022effects,kuhne2024virtual,hung2023facilitators}.
However, one disadvantage of virtual reality is the lack of haptic feedback. 
This lack of tactile information makes interactions potentially less intuitive.
When VR is employed as a training simulator, it is essential that the environment closely replicates real-world conditions to facilitate knowledge transfer.
In contrast, when VR is used to record demonstrations for robot learning, visual fidelity may be less important than the quality and reliability of user behavior during task execution.

\begin{figure}
\centering
\includegraphics[width=0.6\textwidth]{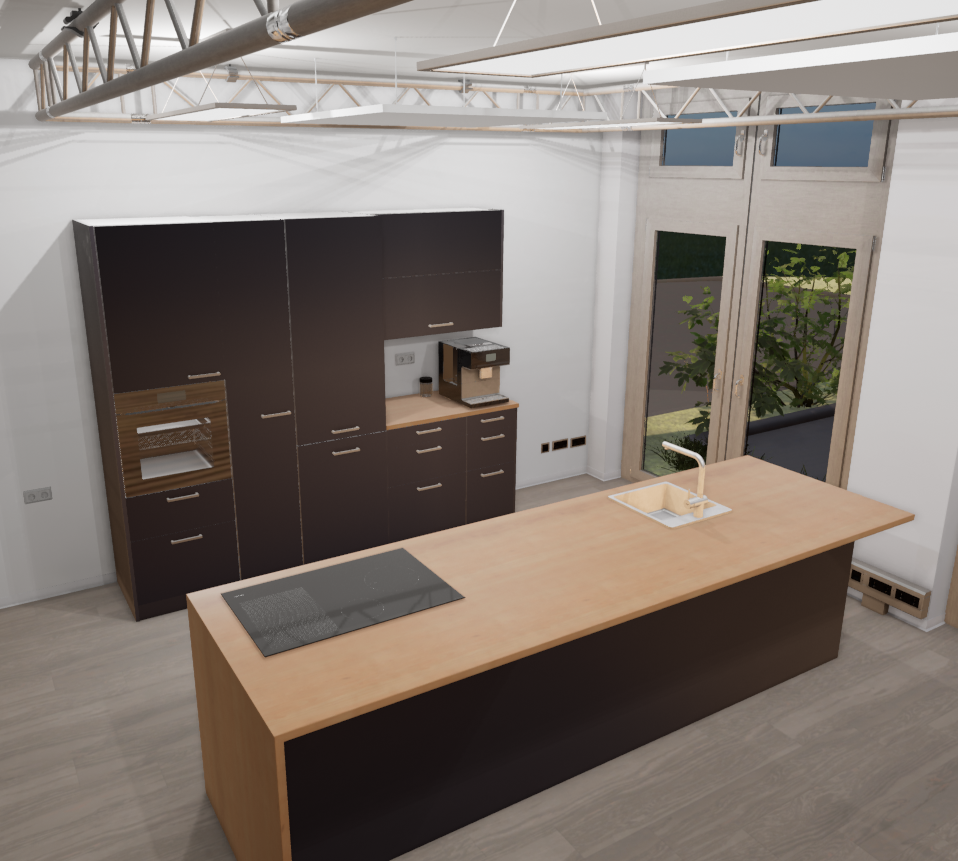}
\caption{The simulated virtual kitchen for the experiments}
    \label{fig:topright}
\end{figure}

\noindent If users are unable to do tasks reliably or perform unnecessary actions, valuable training data may get lost and the learning from the few given shots would be more difficult.
Beyond robotics, analyzing user behavior in VR through trajectory data is also relevant for rehabilitation.
Trajectory segmentation and analyses can help identify movement inefficiencies, compensatory strategies, and deviations from optimal motion paths.
For example, in stroke rehabilitation, patients may exhibit unintended pauses, excessive corrections, or asymmetric motion patterns, which could be detected through trajectory-based analysis.
Trajectory segmentation and analysis can compare the patient's movement patterns to those of healthy individuals and identify deviations that suggest compensation.
Our data-driven approach could make VR-based rehabilitation more effective by having more objective data rather than relying solely on subjective therapist assessments and observations.
Especially in such applications, where acceptance and engagement are essential, the choice of interaction devices as well as their appearance is important.
To understand how the combination of input modality and virtual representation—referred to as input configuration—affects user experience and task-specific interaction behavior in VR, we investigated three configurations across a set of everyday tasks.
Our goal is to identify which input configuration is most suitable for which type of task, depending on whether efficiency or natural interaction is prioritized.
Like categorized in the literature\cite{golinkoff2008toddlers,vollmer2014robots}, task types can be for example manner oriented (e.g. cutting bread) or goal oriented (e.g. setting up the table) and the important aspects can be controlled error free or natural behavior where errors do not weigh as strong.
Therefore, and as a secondary contribution we developed a trajectory analysis to quantify execution quality with respect to spatial similarity and semantics.
The resulting insights could for example support adaptive rehabilitation programs that offer real-time feedback and personalized exercises.

In this work, we refer to each combination of input hardware and its corresponding visualization as an \textbf{Input Configuration (IC)}.
This allows us to systematically study the joint effect of input modality and visual feedback on user performance.
The three Input Configurations used in our study are:
\begin{itemize}
  \item \textbf{M} – Motion capture gloves with articulated hand visualization,
  \item \textbf{H} – Controllers with hand visualization reconstructed from sensor input,
  \item \textbf{C} – Controllers with controller visualization.
\end{itemize}

\noindent These configurations represent different trade-offs between immersion, precision, and realism in VR interactions, and serve as the basis for all subsequent comparisons.
We implemented a VR application in Unreal Engine supporting naturalistic interactions such as grasping, touching, and pointing. 
These ICs were evaluated in a user study in which participants performed common kitchen tasks such as setting a table, cutting bread, and pouring liquids.
To assess the effects of each IC, we examined both subjective and objective measures: 
subjective metrics included usability and workload, while objective metrics focused on trajectory similarity, unnecessary actions, and input errors.
This work contributes to the understanding of how input modality and visualization affect task execution in VR and proposes a trajectory-based framework to analyze interaction quality.

\section{Related Work}
\label{sec:background}

In recent years, the development of VR hardware has seen a significant growth.
Cable-bound devices got replaced by wireless attachments, the tracking accuracy was improved, and hand tracking or motion capture gloves became more popular~\cite{DBLP:journals/tog/HanLCTZPYTAWNDY20,Cox_Hicks_Gopsill_Snider_2023}.
Current Virtual Reality (VR) technology enables precise motion tracking \cite{niehorster2017accuracy,sitole2020application} and interaction simulation, making it a valuable tool for training and movement recording.
However, the choice of input modality and visualization can significantly influence the quality of recorded demonstrations.
For example, displaying hands or controllers can influence immersion and intuitiveness, which in turn affects user behavior \cite{vona2025hands}.

\noindent Prior research has explored VR control methods using virtual kinematic hands with hand tracking~\cite{DBLP:conf/vr/DelrieuWG20} or template matching with controllers~\cite{DBLP:conf/mc/SchaferRS22}.
While hand tracking enhances immersion, controllers offer better control and precision in grasping and typing tasks~\cite{DBLP:conf/qomex/Voigt-AntonsKA020}.
However, most studies focus on simple interactions in constrained environments and to the best of our knowledge, no prior work has systematically compared different input configurations with respect to both user experience and task execution behavior in realistic ADL multi-step tasks.
Although various VR control methods and visualizations have been evaluated for usability~\cite{DBLP:conf/hci/JohnsonFPEW23,vona2025hands,DBLP:conf/qomex/Voigt-AntonsKA020} and speed \cite{DBLP:conf/vr/ArgelaguetHTL16,DBLP:journals/mti/KangasKMJR22} their impact on the generated data quality remains an open question.

\noindent Since research has shown that users have varying preferences regarding feedback, controls, and visualization \cite{vona2025hands}, flexible controls  with an underlying recording framework are necessary for our approach.

\noindent One system for recording everyday activities is AM-EvA~\cite{beetz2010towards}.
It captures detailed motion data, segments it and categorizes the segments by actions to synthesize execution plans for everyday tasks.
AM-EvA is based on KnowRob~\cite{DBLP:conf/iros/TenorthB09,DBLP:journals/ki/TenorthJB10,DBLP:conf/icra/BeetzBHPBB18} which is a reasoning engine for episodic memories.
In 2019 VR technology was incorporated to simulate human environments and recording manipulation activities within them~\cite{DBLP:conf/icra/HaiduB19}.
However, these systems focus on structured, expert-driven demonstrations and lack realism in everyday tasks. 
While they provide valuable insights for robot learning, their application to broader fields such as rehabilitation and user-centered VR training remains limited.
There exist many studies and surveys, which argument how to use VR for rehabilitation training for example for children with cerebral palsy \cite{warnier2020effect,liu2022effects,chen2018effectiveness} or upper limb rehabilitation to relearn daily tasks\cite{levin2015emergence}.

These studies using VR for rehabilitation show promises in rehabilitation, mostly because the patients repeated the tasks more often than without the use of VR.
However, many systems need personell to operate the systems and to set up the configurations and can only give results based on observations.
Additionally, most VR applications rely on built-in controllers, which can be difficult to handle for some users.
Since state-of-the-art devices like the Meta Quest 3 include built-in hand tracking \cite{metaquest3}, switching between input modalities is now more feasible.
The question when this is necessary, or useful is what we want to answer. 
To do that, the different controls and their effects need to be analyzed precisely, for example by comparing the captured movements, called trajectories.

For comparing traditional trajectory distance measures, such as Frechet Distance~\cite{eiter1994computing,DBLP:conf/edbt/TangYMW17} and Dynamic Time Warping~\cite{sakoe1978dynamic,kassidas1998synchronization} can be used as similarity measures.
However, they only assess the entire movement paths while ignoring any other data and interaction trajectories in VR involve more information than only movements. 
Interaction trajectories include position, pose, and event data (e.g., grasp or interaction events), all of which contribute to the quality of the execution, making it more complex to evaluate.
Using the semantic information of the execution leads to the concept of semantic trajectories~\cite{bogorny2014constant,parent2013semantic,yan2013semantic}, which integrate contextual information like actions or events.
Semantic trajectories are often used for geospatial data, where the semantic information can be the locations for taxistops or similar\cite{yan2013semantic}.
There also exist work on similarity search on semantic trajectories \cite{chen2024efficient,shang2024efficient}, however, how this can be used to measure the quality of those trajectories is not yet explored.
We already used basic semantic information and have shown that a similarity measurement can be improved by segmenting the trajectory \cite{helmert2025semantic}.

In this work we put the focus on assessing the impact of different input devices and visualizations in VR on the demonstration recordings as well as the usability and the workload for the user based on different tasks.
By leveraging trajectory analysis, we combine different metrics to evaluate interaction quality and provide insights into how the control modalities and their visualizations influence performance for different task types. 
These findings might help determine the most effective IC for various VR-based tasks.


\section{Research Questions and Study Objectives}
\label{sec:research_questions}

\noindent Therefore, in this paper we aim to answer the following research questions:

\begin{itemize}
    \item \textbf{RQ1:} Which input configuration provides the best user experience in terms of usability and workload?
    \item \textbf{RQ2:} How does the choice of input modality and its visualization influence user behavior and task execution, particularly in relation to the recorded trajectories?
\end{itemize}

\section{Study Design}
\label{sec:study_design}
To assess the impact of different control visualizations in everyday kitchen tasks, we implemented the described ICs.
They offer the same set of possible interactions of grasping, moving objects, and doing the pointing or rating interaction but differ in how these interactions are triggered.
Grasp is the most common interaction in VR and can be used to pick up and move objects or opening drawers and containers like the fridge.
While carried, objects can be reoriented or thrown away like it can be done in reality, ensuring intuitive object handling.
Grasping and moving objects has to be used in nearly any task to complete it.
Another interaction is pointing and rating. 
Since pointing and rating can be powerful tools to deliver specific information without the use of words we also used them in our experiments.

\subsection{Input Configurations}
\label{subsec:study_controls}
\begin{figure}
    \centering
    \includegraphics[width=\linewidth]{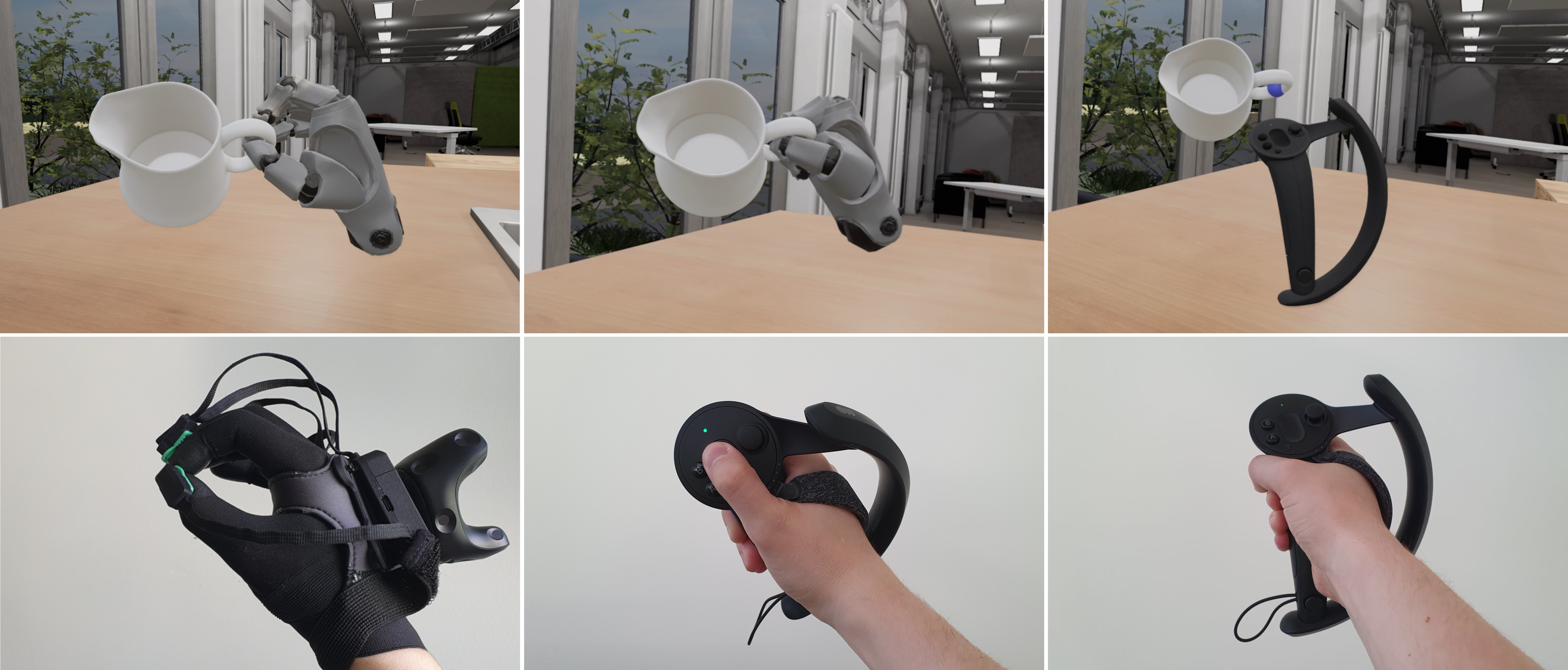}
    \caption{The ICs and their visualization. The lower row shows the condition which are either the Manus motion capture gloves (left column) or the Valve Index controller (middle and right column). The upper row shows the virtual visualization which is either the hand reconstruction (left and middle column) or only the controller (right column). For all images currently the virtual milk pitcher of the pouring scenario is grasped and held.}
    \label{fig:controls}
\end{figure}

\paragraph*{M - Manus Hands}
The first IC used the Manus motion capture hands and input device. 
It was visualized in VR with an accurate virtual hands that mimic the motions of the real one and is shown in the leftmost column of Figure \ref{fig:controls}.
Since the level of immersion is defined of how accurate the real world is simulated\cite{slater1994body,slater2009place} motion capture gloves have the highest immersion among the used controls and enable intuitive  gestures.
To have grasping as natural as possible, we also added more than one grasping trigger.
In reality, for grasping a knife, or a handle the hand is often nearly a fist, while grasping a small objects is done by a pinch.
Therefore fist as well pinching was considered grasping and any object could be grasped by any of these gestures. 
Like for grasping, natural gestures were used for rating as well as pointing.
Rating could be done with a thumbs up or down gesture.
A thumbs up would give a positive rating and thumbs down gives a negative rating.
For the pointing gesture, the hand must be closed with the index finger spread out. 
While doing this gesture a laser pointer was emitted from the index finger with a target marker at the end. 
The direction hereby was taken from the hand bone direction to counteract jitter which would occur when using index finger.
When the gesture was resolved, the target marker froze at the position showing a thumbs up or thumbs down image.
Now participants could use the thumbs up or down rating gesture for depending if they were pleased with the target pointed at or not.

\paragraph*{H - Valve Index Controller with Hands}
The second IC used the Valve Index Controller, as it is a technical advanced controller with good usability.
It is not unpleasant to wear and has all the standard inputs like joystick, buttons, trackpad and trigger.
In Virtual Reality, the controller was not displayed.
Instead the real hand was simulated using the touch and pressure sensors of the controller.
This IC is shown in the center column of Figure \ref{fig:controls} and combines usability with natural interactions, while having a high immersion through reconstructing hand poses from the touch data of the Index Controller \cite{valveindexcontrollers}.
We believe the fixed grip design of the Valve Index controllers with its ergonomic design improves usability in comparison to other state of the art controller.
Grasping with the controllers can be done by either using the trigger button, as it is done in most games, or by squeezing the controller which is measured by the pressure sensor indicating how tight the controller is held.
This way it is ensured that most anticipations of the grasping controls can be satisfied.
The rating was done by tilting the joystick.
Upwards would be thumb up and downwards thumb down.
For pointing, any of the buttons on the controller could be pressed. 
While any of the buttons pressed, like for first IC, a laser with a target marker at the end was emitted from the controller in forward direction, indicating the pointing position.
When the button was released, the laser target marker froze and the laser was turned off.
Now user could use the rating interaction to state if they liked their pointing target or not and redo it if wanted.

\paragraph*{C - Valve Index Controller without Hands}
The third combination uses the same controller as the second, but instead of the hand is here the controller visualized and the hand is made invisible as shown in the right column.
Thus the participants can see the controller and where each button is.
This IC is shown in the right column of Figure \ref{fig:controls}.
Interactions with this IC are done at a highlighted blue sphere at the top of of the controller.
In the figure this sphere can be seen at the handle of the milk pitcher.
Showing only the Index controller itself supports realistic visuals and better coherence between real and virtual world at the cost for immersion.
Besides that the interactions were done at the blue sphere instead of the hand, this IC shares the same interactions and interaction trigger as the H input configuration.

\subsection{Tasks}
\label{subsec:study_tasks}

We then designed a set of distinct tasks, each with different solution strategies.
Each task can be solved with any IC using the available interactions.

\paragraph*{Introduction}

\begin{figure}[t]
\centering
\begin{subfigure}[t]{0.48\textwidth}
    \includegraphics[width=\linewidth]{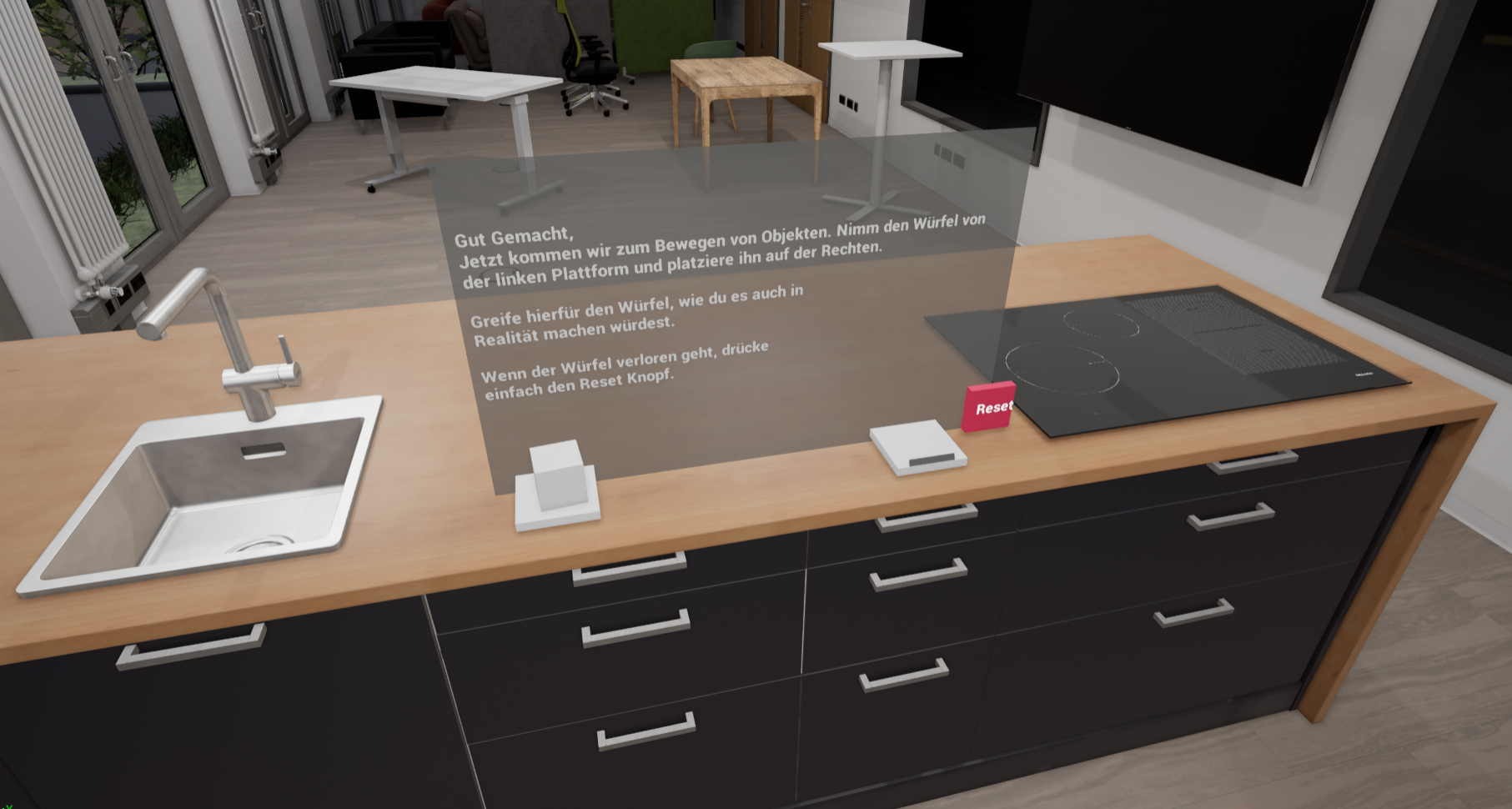}
    \caption{Introduction scene. The current task introduces grasping and requires the white block to be placed on the right platform.}
    \label{fig:sc:intro}
\end{subfigure}
\hfill
\begin{subfigure}[t]{0.48\textwidth}
    \includegraphics[width=\linewidth]{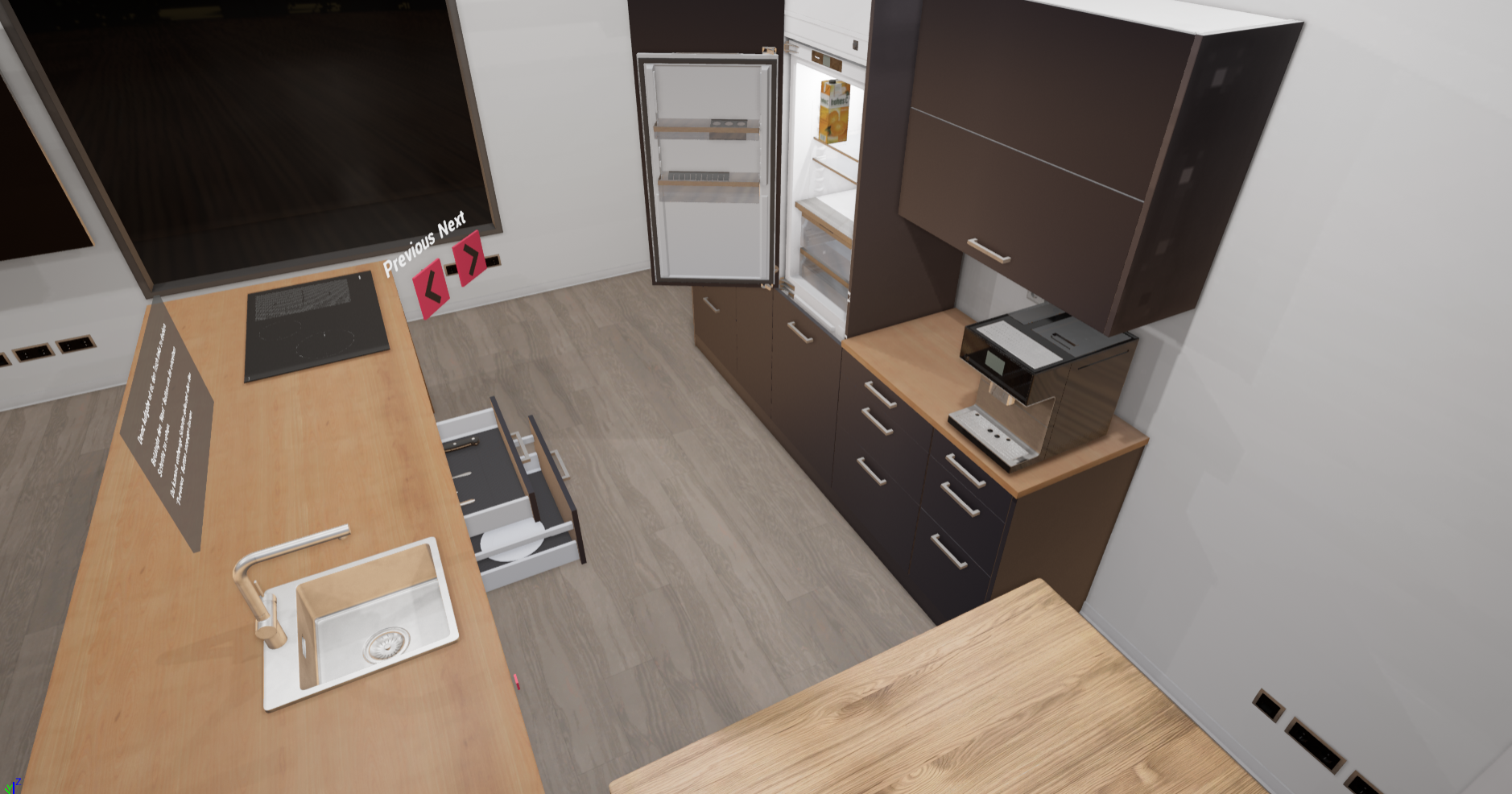}
    \caption{Table Setup scene with open drawers and fridge. In the initial state, all storage elements are closed.}
    \label{fig:sc:table}
\end{subfigure}
\caption{Overview of two scenes used in the study: (a) Introduction and (b) Table Setup.}
\label{fig:scenes}
\end{figure}

While the \textit{Introduction} is the first task every participant has to solve, it is neither recorded nor did we evaluate anything in this task.
It interactively introduces the possible controls and interactions to the participants and is visualized in Figure \ref{fig:sc:intro}. 
By doing so it offers a small and easy task for each interaction where users have to complete it to progress.
This way we prevented overreading the instructions and ensured all interactions were done at least once before the real tasks begin.

\paragraph*{Table Setup Task}

The \textit{Table Setup Task}, illustrated in Figure \ref{fig:sc:table}, incorporates the first assignment and is considering some basic pick and place operations.
Users should take items out of the drawer and the fridge and place it on a table.
Those items are a dessert spoon, a fork, a knife, a plate, and an orange juice carton.

\paragraph*{Dishwasher Task}

\begin{figure}[t]
\centering
\begin{subfigure}[t]{0.48\textwidth}
    \includegraphics[width=\linewidth]{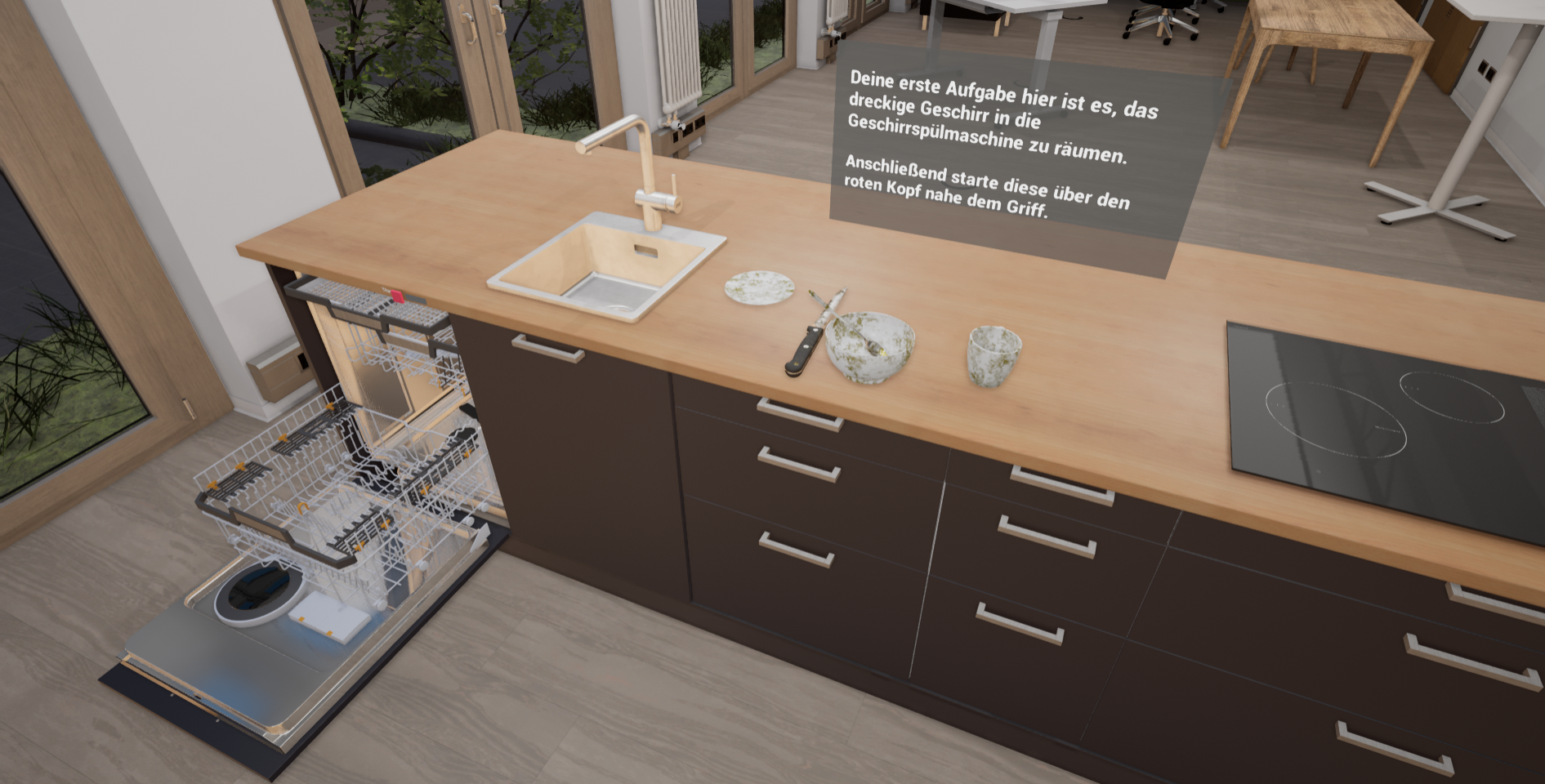}
    \caption{Dishwasher scene. Participants must place the dirty dishes in the dishwasher (initially closed) and start it afterward. Finally, they should empty it and place the dishes back into the marked drawer.}
    \label{fig:sc:dishwasher}
\end{subfigure}
\hfill
\begin{subfigure}[t]{0.48\textwidth}
    \includegraphics[width=\linewidth]{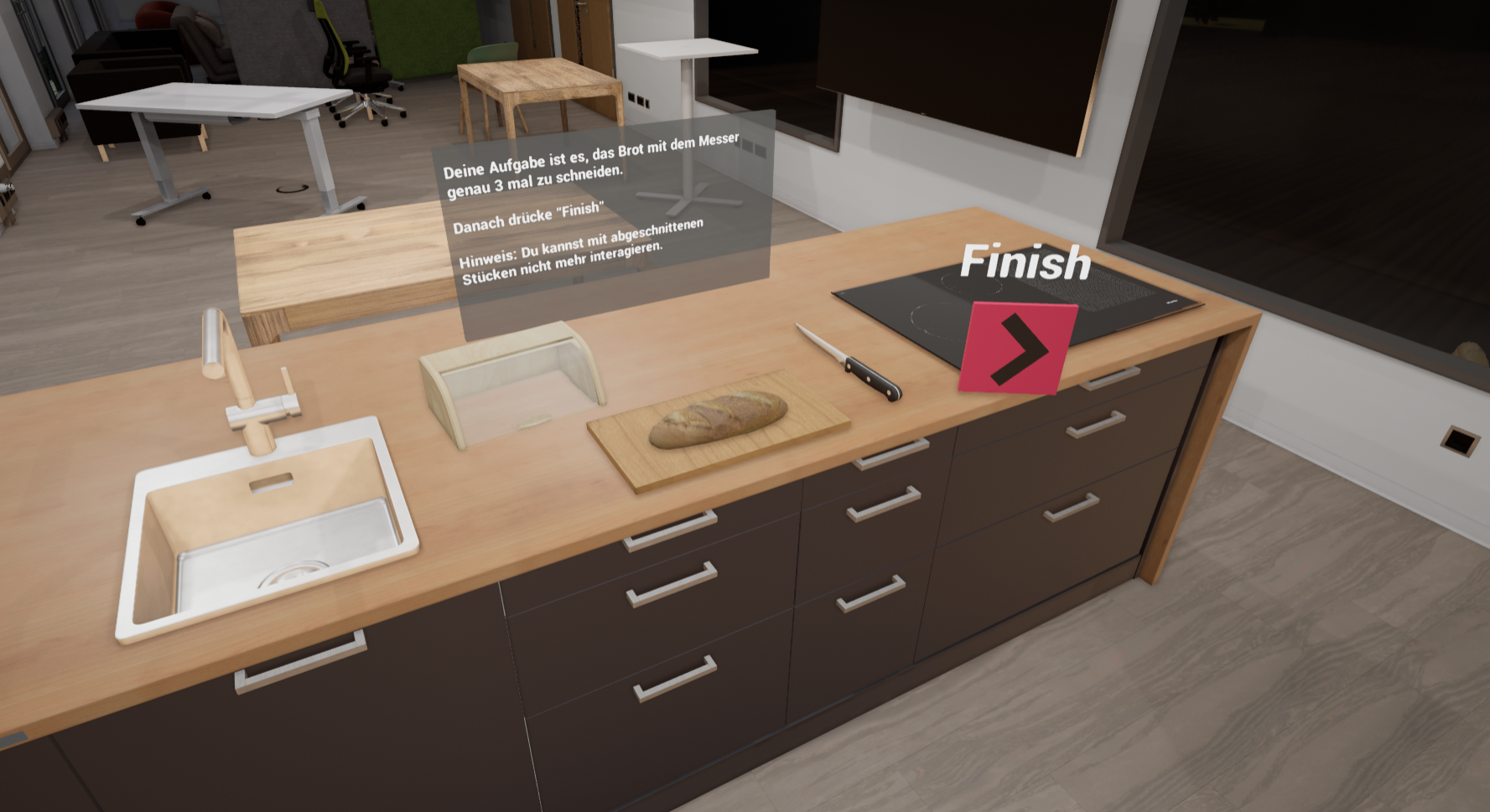}
    \caption{Cutting scene. Users are asked to pick up the knife and cut the bread three times.}
    \label{fig:sc:cutting}
\end{subfigure}
\caption{Two scenes used in the study: (a) Dishwasher task focusing on precision and placement, (b) Cutting task focussing of the motions of the knife.}
\label{fig:scenes2}
\end{figure}

The \textit{Dishwasher Task} is displayed in Figure \ref{fig:sc:dishwasher} and considers filling the dishwasher with dirty dishes, starting it, and then emptying it, placing all the dishes back in highlighted drawers.
It also focuses on pick and place tasks but emphasize more on precision, since it demands to place the dishes precisely into the dishwasher.

\paragraph*{Cutting Task}

\textit{The Cutting Task} changes the focus and is displayed in Figure \ref{fig:sc:cutting}.
As the first manner-oriented task, users were asked to cut a bread with a given knife with three cuts.
The knife was initially placed right beside the bread and users should pick it up, then slice the bread and then put it back down.

\paragraph*{Cleaning Task}

\begin{figure}[t]
\centering
\begin{subfigure}[t]{0.48\textwidth}
    \includegraphics[width=\linewidth]{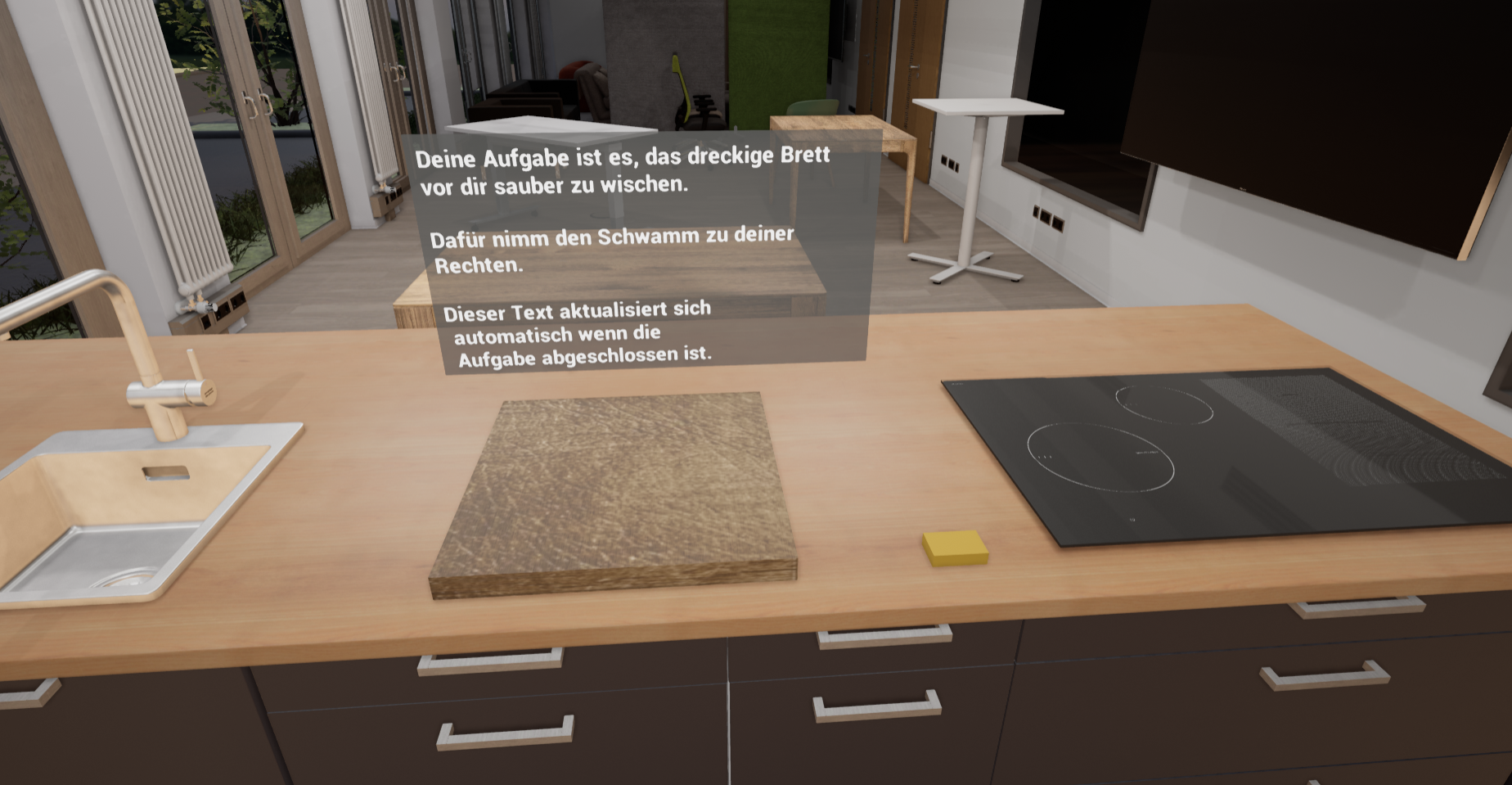}
    \caption{Cleaning scene. Users should pick up a sponge and then clean the board in the middle.}
    \label{fig:sc:cleaning}
\end{subfigure}
\hfill
\begin{subfigure}[t]{0.48\textwidth}
    \includegraphics[width=\linewidth]{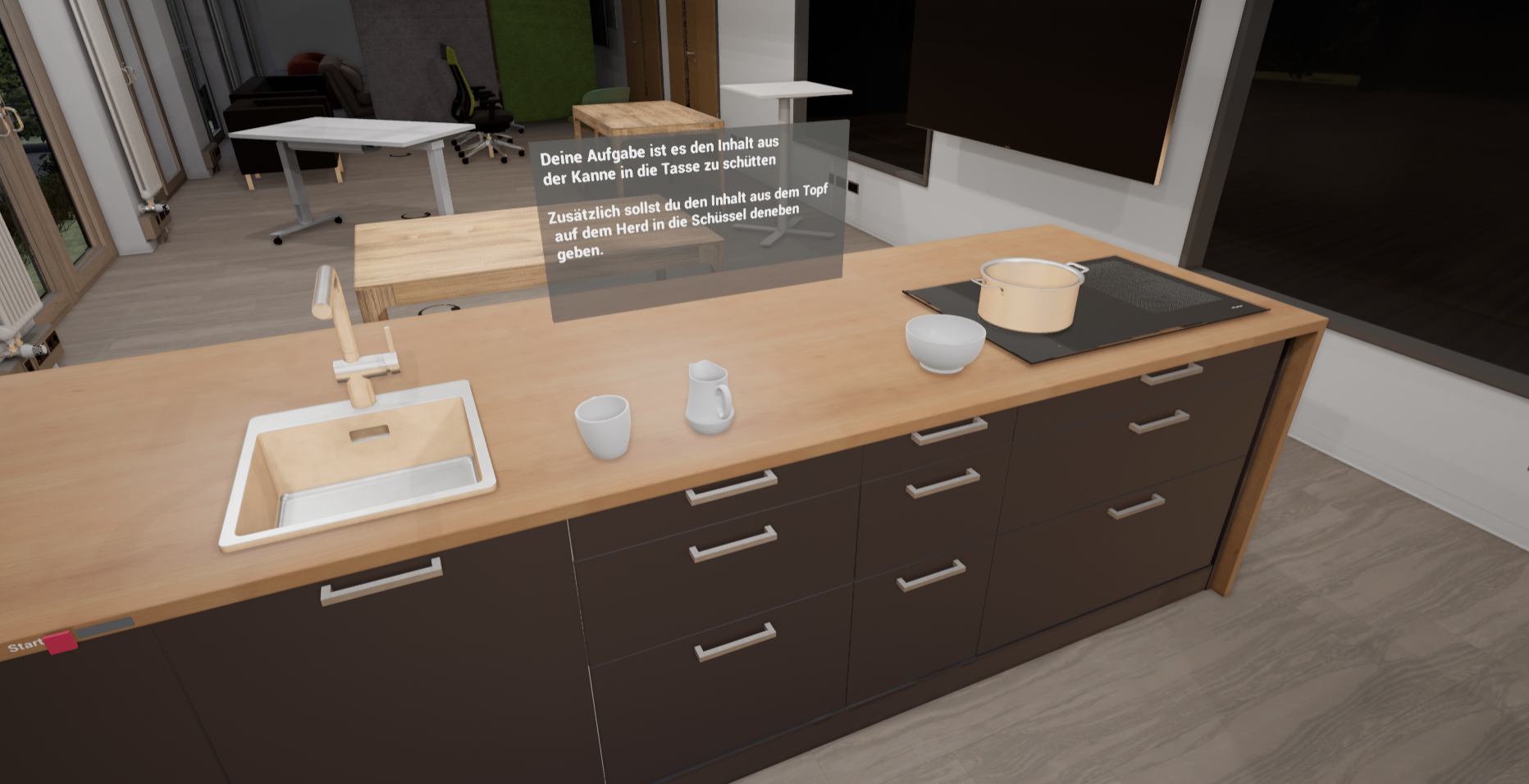}
    \caption{Pouring scene. Users should fill the contents of a pitcher into a cup and the contents of a pot into a bowl.}
    \label{fig:sc:pouring}
\end{subfigure}
\caption{2 Scenes focused on continuous motion: (a) Cleaning, requiring vertical movement control; (b) Pouring, requiring slow and precise execution to avoid spilling.}
\label{fig:scenes3}
\end{figure}

The same manner-oriented focus had the \textit{Cleaning Task} shown in Figure \ref{fig:sc:cleaning}. 
Here the participants could pick up a sponge, clean a defined surface and then place it back down.
The initial and end position of the sponge does not matter as much as the way it was moved during the cleaning operation.

\paragraph*{Pouring Task}

\noindent The \textit{Pouring Task} which is illustrated in Figure \ref{fig:sc:pouring} is also manner oriented. 
It considers containers containing liquids, simulated with small spheres that should be filled into other containers.
The contents of a milk pitcher should be filled into a cup and the contents of a pot should be filled into a bowl. 
Similar to the Cutting and the Cleaning Task, the way the interactions is done is more important than the correct start and end location of the used container.
However, not only the paths but also the execution speed is important.
In this task, slow and precise movements are necessary to prevent spilling.

\paragraph*{Pointing and Rating Task}

\begin{figure}
\includegraphics[width=0.48\textwidth]{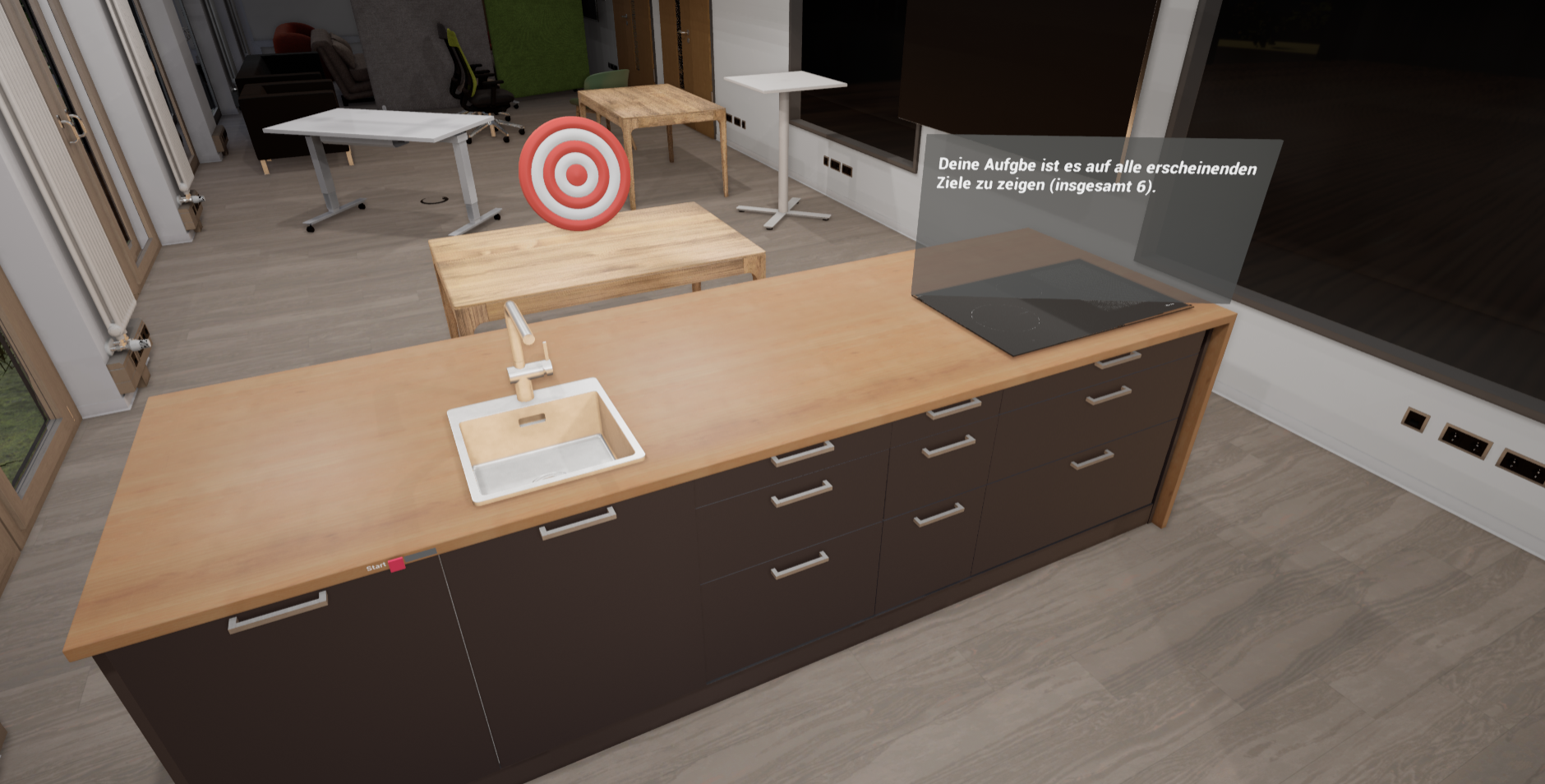}
\caption{Pointing and Rating task. It shows the situation where the user shall point on the displayed target using either the pointing gesture or buttons depending on the IC.}
    \label{fig:sc:pointing}
\end{figure}

The last exercise implemented was the Pointing and Rating Task.
For a comprehensive comparison of the ICs, we also implemented and evaluated these tasks involving gesture-based actions.
Participants here were not ask to interact with objects, but instead should trigger some interactions. 
Users had to give some positive and negative ratings and then point at six different targets, each distinct but at fixed positions in the virtual room which can be seen in Figure \ref{fig:sc:pointing}.
In this scene, the users' ability to remember the controls or gestures as well as the precision of the pointing action is recorded.
Our target was to evaluate how users could remember the gesture invocations of the different controls and how accurate objects in variable distances can be pointed at.

\subsection{Implementation and Data Recording}
\label{subsec:implementation}
The scenes and interactions were implemented in \textit{Unreal Engine 4.27} in a fork\footnote{\url{https://github.com/RBeierling/RobCoG}} of RobCoG\footnote{\url{https://github.com/robcog-iai/RobCoG}}.
As described, we selected a kitchen as our virtual environment.
The given model serves as a digital twin of a real kitchen which is located at the University of Bremen.
All tasks outlined in this study are implemented in this virtual kitchen which is shown in Figure \ref{fig:topright}.
The recordings were done in the NEEM format~\cite{DBLP:conf/icra/BeetzBHPBB18}.
NEEM stands for \textit{Narrative Enabled Episodic Memory}.
This format is well-established for storing and recalling past events in robots \cite{DBLP:conf/ecai/AhrensPBFBH23,DBLP:journals/ki/BartelsBB19,DBLP:conf/iros/TenorthB09}.
NEEMS themselves contain structured and detailed records of time-indexed experiences that combine symbolic and sub-symbolic data to create comprehensive narratives of actions and interactions~\cite{beetz2020neem}.
Since a NEEM holds time-series data of the poses for every object in a scene, the task execution can be replayed afterward to apply additional quality checking and assessment methods.

\section{Methods}
\label{sec:evaluation_methods}

Our approach utilizes virtual reality to accurately record positions and interactions directly from a virtual scene, eliminating the need for a video-based motion capture setup.
The virtual scene constitutes the digital twin of a real kitchen that features a fully interactive interior which can be seen in Figure \ref{fig:topright}.
Thereby we achieve that most actions that are possible in the real kitchen can be replicated in the virtual scene.
To answer the research questions, the following measurements were used.

\subsection{Research Question 1}
\label{subsec:data_collection}

To address the first RQ, user experience is evaluated based on usability and workload.
We assess usability using the System Usability Scale (SUS)~\cite{brooke1996sus} and workload using NASA Task Load Index (NASA-TLX)~\cite{HART1988139}.
Additionally, participants provided qualitative feedback on what they liked, disliked, and suggested for improvement.

\subsection{Research Question 2}

To address the second RQ, we analyze the influence of input modalities and visualizations on recorded task executions.
Key metrics include unnecessary actions, execution time, and interaction precision. 
Further, we recorded the movements of the user and all objects within the scene.
This was achieved by logging the position, orientation, and interaction states per object at 20 Hz throughout the experiment.
By integrating both subjective and objective evaluations, we aim to gain a better understanding of how different control methods influence user behavior, interaction efficiency, and task execution quality.
Frequent grasping and releasing, for example, might indicate user uncertainty or problems with the control.
Also, considering robot training, a robot that learns from a demonstration with consistently too many grasps might consider all grasps as crucial and this would ultimately result in a sub-optimal learning and task execution afterwards.
To pinpoint exactly where problems or inaccuracies are, we segment the recorded trajectories into sub-trajectories and then compare those sub-trajectories with the same goal.
A trajectory in our study represents the action and motion path of an object or hand during a task.
It consists of time-indexed points that store position, orientation, and actions (e.g., "grasp", "cut") as well as the semantic information like grasping a knife or cutting bread.
Since each task consists of multiple phases (e.g., reaching, cutting, releasing), it is reasonable to split the trajectories at the change of these phase into subtrajectories and then analyze the specific subgoal.
Our trajectory definition is adapted from \cite{bogorny2014constant} which focuses on the definition semantic trajectories and subtrajectories for example ones that were extracted from GPS data where different points of interest were visited.
For our case we adapted their definition to fit our case of movement trajectories.
Given a set of trajectories $T$ each belonging to an object, a trajectory $tr \in T$ is  defined as $$ tr:= (tid, oid, S, g)$$
where $tid$ is the trajectory id, the $oid$ is the object id, $S$ is a list of all subtrajectories, and $g$ is the overarching goal of the trajectory. In the case for the task of cutting the bread bread three times it would be \textit{g:= "Cut the bread 3 times"}.
A subtrajectory $st \subset tr$ is defined as $$ st:= (sid,tid,P,tstart,tend,g_s)$$ where
$sid$ is the id of the subtrajectory, $tid$ is the id of the respective main trajectory, $P$ is the list of all semantic points in that trajectory, and $tstart$ and $tend$ denote the start and the end timestamp of the subtrajectory.
We added the parameter $g_s$ as subgoal to the original definition as task mostly have implicit subgoals like \textit{grasp the knife} or \textit{cut the bread}.
A point in a subtrajectory $pt \subset st$ is defined as $$pt:=(pid,sid,pose,t,A)$$ where
$pid$ is the id of the point, $sid$ is the id of the respective subtrajectory, $pose$ is the pose of the object specified by the \textit{oid} at that specific time, and contains the position and the orientation of that object defined by the engine origin, $t$ is the timestamp for the recorded point, and $A$ is a list of actions that were observed at that point in time.
$A$ can be empty or contain an arbitrary amount of actions.
For the knife it might contain grasping and cutting at the same time.
These definitions suit all the trajectories we can record in VR and enable an automatic segmentation stated above to distinguish between different parts of the task execution without the need of sophisticated matching algorithms.

The task description infers that the bread has to be cut three times.
Therefore the optimal trajectory considering only the semantic aspects would start where no action is currently done, followed by a grasp of the knife which in turn is followed by three cuts and afterward the placing down of the knife leaving with and empty action set again.

\subsubsection{Trajectory Segmentation and Analysis}
\label{subsec:trajectory_analysis}

Since the complete task of cutting a bread with a knife consists of multiple parts which each can be of a good or a a bad quality on its own, it would not be applicable to consider one measurement for the whole trajectory.
Especially since for most tasks the respective subtasks changes from time to time it is necessary evaluate each part separately.
Therefore, we split the trajectory into different segments or subtrajectories which are then evaluated individually.
For example is the focus of the trajectory where the hand approaches the knife different than the part of the trajectory that actually cuts the knife.
The target of the first is having the correct endpoint, i.e. the knife in order to grasp it, where the latter focuses on the cutting movements during the actions.
Therefore, we chose as segmentation points, the timestamp where the subtask changes.
For example, at the time the knife is grasped, the subtask of grasping the knife changes to move the knife to the bread where the cut is to be made where the subtask in turn changes to actually cut the bread.

\subsubsection{Semantic Trajectory}
To analyze the semantics, we focus the actions occurring within a trajectory.
With this the duration of actions, the position where they were happening and additionally the correct order and amount of interactions can be confirmed.
For example to confirm the amount and order of subtasks for the knife in the task, with the definition of the trajectory from above considering the only subtrajectory being the trajectory itself we remove all points from the trajectory where the previous point shares the same set of actions $A$.
With the set of legal action labels $\{grasp,cut\}$ for cutting the ideal sequence of points for a task where a user is required to grasp a knife and perform a single cut on an object would then be resembled by the following trajectory
\begin{align*}
P &= ((p_1, s_1, \text{pose}_1, t_1, \{\}),\\
&\quad \enspace\;(p_2, s_1, \text{pose}_2, t_2, \{\text{grasp}\}),\\
&\quad \enspace\;(p_3, s_1, \text{pose}_3, t_3, \{\text{grasp}, \text{cut}\}),\\
&\quad \enspace\;(p_4, s_1, \text{pose}_4, t_4, \{\text{grasp}\}),\\
&\quad \enspace\;(p_5, s_1, \text{pose}_5, t_5, \{\})\bigr).
\end{align*}

With $t_5 - t_2$, the time from the first interaction to the end of the last interaction can be calculated and therefore represents the duration of the grasp interaction with this item.
The semantic information can be shortened to $$\{\},\{grasp\},\{grasp,cut\},\{grasp\},\{\}$$ if the timings and additional information are not important. 
It gives us the pure semantic trajectory.
While achieving a perfect spatial trajectory is nearly impossible, achieving the perfect semantic trajectory can be done. 
When the execution of a participants fits such a perfect semantic trajectory it means that the participant did no unnecessary interactions and the order of interactions is correct.
This is one crucial aspect for high quality demonstrations.
Moreover, these semantic trajectories can also be used to automatically find these unnecessary interactions and therefore can be used to make changes in the environment that could fix those.
They therefore can be used as a foundation for a framework that automatically adapts to imperfect movements with the aim to improve those.
Additionally, showing that the different ICs result in different semantic trajectories can show additional strength and weaknesses of the ICs.

\subsubsection{Errors and Interaction Times}\label{sec:Methods:Tasks}
To calculate the deviation from an optimal semantic trajectory we uses the Levenshtein distance \cite{levenshtein1966binary} as measurement.
Originally meant to state the difference between two strings by calculating the necessary deletions, insert and changes, it can also be used to calculate a distance between two semantic trajectories.

With the Levenshtein distance calculating the difference to an optimal semantic trajectory it is used as an indicator of the amount unnecessary, missing and wrong actions.
Whether unnecessary actions should be counted as error or not depends on the viewpoint, if we focus on optimal executions like for robot training, it should be counted as such, but if we focus on analyzing the user behavior it should be treated separately.
Some participants might take extra steps due to personal preference, hesitation, or exploration, which may not necessarily be a mistake but rather a reflection of their interaction style.

Therefore we define errors as actions that made solving the task unsuccessful. 
For the cutting tasks this would be for example if the cutting amount is not equal three or for other tasks if specific objects that had to be used were not used at all or in a wrong way.
The latter can be found for example by examining the amount of grasps.
For most tasks there exists an optimal amount of grasps necessary to solve the task. 
Having less than that implies the tasks was not completed successfully while having more implies there were unnecessary actions.

But even if the task is well done and the Levenshtein distance equal zero, there might be differences, for example in different times and durations.
Therefore, we used object times, which indicate how long an object was interacted with, idle times which indicates how long no action was present and lastly, task times which measure how fast the participants could solve the tasks.
The task time does not include reading the task description in the beginning.
Instead the time is taken from the moment the first action started till the last action finished. 
Thus, it can be calculated automatically from the trajectories without any manual annotation, minimizing human errors.

For the Pointing and Rating Task also the precision was evaluated.
It was used to measure how the ICs differ in accuracy.
Here, the precision was measured as the distance between the point where the target was hit and the center of the target object.

\section{Study Procedure}
\label{sec:study_conduction}

The study was conducted with 60 participants (32 male, 27 female, 1 diverse)  which where acquired through flyers, and held in the XR-Lab of the Interactive Robotics in medicine and care workgroup of the Bielefeld University.
Besides that all participants had to be at leas 18 years old, there were no exclusion criteria.
They mostly came from social sciences (n = 21), engineering (n = 17), and education (n = 7) and were between 18 and 34 years old (n = 57).
Most of the participants reported to use VR less than one time a month (n = 52).
Only 7 participants stated they had a very low or low technical affinity while most participants had a medium to very high technical affinity (n = 53).

The study itself employed a between-subjects design.
Each session started with a brief introduction to the topic, emphasizing that the tasks performance will be evaluated.
Participants then were randomly and counterbalanced assigned to one of the three control visualization combinations without knowing of the others.
Additionally, besides the introduction and the Pointing an Rating tasks, the order of the tasks was predefined pseudo randomly.
The tutorial and the introduction was the first scene for all participants to ensure a basic understanding of their IC. 
The scene considering pointing and rating was always presented last, so we could evaluate whether a participant could remember the rating and pointing controls.
Since that time between learning in the introduction and applying the knowledge in the Pointing and Rating Task matters, this task was fixed at last.
After completing all tasks, the participants filled out the SUS and TLX questionnaires.
They were asked to give pro and contra arguments of their experience and could then add additional comments as well.

\section{Results and Analysis}
\label{sec:results}

\subsection{User Experience and Workload (RQ1)}
\label{subsec:results_user_experience}
\begin{table*}[t]
    \vspace{0.2cm}
    \centering
    \begin{adjustbox}{max width=\textwidth, center}
    \begin{tabular}{l|c|c|c|c|c|c}
         & \textbf{SUS} & \textbf{Raw TLX} & \makecell[c]{\textbf{TLX}\\\textbf{Mental Demand}}
 & \makecell[c]{\textbf{TLX}\\\textbf{Physical Demand}}
 & \makecell[c]{\textbf{TLX}\\\textbf{Performance}}
 & \makecell[c]{\textbf{TLX}\\\textbf{Frustration}}
 \\
        \hline
        \textbf{M} & 82.5 (5.68) & 16.13 (10.96) & 13.5 (15.14) & 12.25 (12.82) & 31.75 (20.08) &  18 (19.83) \\
        \textbf{H} & 77.63 (16.95) & 17.92 (15.96) & 24.25 (26.72) & 11.5 (13.48) & 24.25 (18.30) &  22 (24.94) \\
        \textbf{C} & 84.00 (5.47) & 11.83 (5.40) & 17.75 (18.74) & 8 (7.85) & 19.25 (11.27) & 13 (14.18) \\
    \end{tabular}
    \end{adjustbox}
    \vspace{0.1cm}
    \caption{Comparison of the controls motion capture hands (M), controller with hand visualization (H), and controller with controller visualization (C) by mean ($\Bar{x}$) and standard deviation ($\sigma$). All scales range from 0 to 100~\cite{brooke1996sus,HART1988139}}.
    \label{tab:comparison}
\end{table*}

We conducted the Kruskal Wallis significance test to compare the SUS~\cite{brooke1996sus} and the Raw-TLX~\cite{hart2006} of the three control conditions but no significances were indicated.
Additionally, none of the scores of the TLX subquestions showed any significances.
Mean, standard deviations of SUS, TLX, most of its subquestions for the three controls motion capture hands (M), controller with hand visualization (H) and controller with controller visualization (C) are shown in Table \ref{tab:comparison}.
Since each participant only used one of the controls and did not come in touch with the other options, asking for their favorite control was not possible.
The general feedback was mostly positive, independent on the control condition they had.
The only negative feedback was that some users found the controller too bulky for smaller hands such that they could not reach each button easily.

In addition to the user feedback we also want to point out observations made by the supervisor or participants during the sessions. 
The \textbf{Motion capture gloves (M)} suffered occasionally from the grasping intuitions when participants made a fist to grasp items or handle as they thought it would match the grasping gesture.
In the \textbf{Controller with hand visualization (H)} condition, participants tended to have problems grasping small objects (n = 9) and they often did not find the controllers' physical buttons initially (n = 5).\\
Using the \textbf{Controller with controller visualization (C)}, some participants struggled to understand how to use the controllers correctly, attempting to lift objects by pressing the controllers from both sides against the item, rather than using the intended buttons for object interaction (n = 4). 

Both controller based controls had problems with holding items for a longer time. For example users accidentally released grasped objects like the knife (n $>$ 5).

Additionally, some users for these controls forgot how to perform actions such as pointing or rating in the final scene completely (n = 6).
Here we did not count the situations where they could rediscover how to perform these interactions; which happened frequently for these controls (n $>$ 10).

Finally, users of all control types experienced some misinteractions like accidental opening or closing of drawers or doors.

\subsection{Task-Specific Performance Analysis (RQ2)}
\label{subsec:results_task_performance}
Since each task had different objectives, the measurements also varied
In the following the measurements of each tasks are explained and the results summarized in a table.
Additional significance tests are also highlighted.
In the following, and in the tables, \textbf{M} stands for Manus motion capture gloves, \textbf{H} stands for controller usage with hand visualization and \textbf{C} stands for controller visualization where only the controller was shown to the participants.
The fields below these identifiers in the tables describe the median value across all participants for that control and the metric specified.
Since normality and homogeneity of variance did not hold for any of the measures, we did a Kruskal Wallis Test for each measurement.
If any are found (i.e., $p<0.05$), we did the pairwise conover test with holm correction for unparametrized data to check for significances.
Also each p-value is paired with the effect size calculated with cliffs delta~\cite{cliffdelta}.
Finally the significances and their directions are shown in the result column of each table.
Each Table starts with general comparisons like task time, idle time, and grasp count.
The task time is the time it took to complete the whole task and the idle time describes the time in seconds, the user was not interacting actively with the environment during the task.
The grasp count describes the amount of grasped used to solve the tasks. 
Additionally, hand balance was evaluated for tasks where multiple objects needed to be moved. 
Here a value of $1$ means that both hands were used equally often and $0$ means only one hand was used.

In addition to these general measurements each object of the respective task was also analyzed separately regarding grasp time and trajectory similarity. 
For the latter we used two measurements to determine the distance between two trajectories, the discrete Frechet Distance (DFD) and the Dynamic Time Warping distance (DTW).
To get a comparable value for each object we calculated distances to all other trajectories of that object within the same control.
Then we calculated the median for these distances (i.e., the median distance to all other trajectories of that item within the same control).
Therefore object, has two distance values, a DFD and a DTW one, which represent the median distance to the others in the same group according to the metric.
These values represent the similarity of the trajectories within that group for the specific item.
We also calculated the standard deviation to measure the similarity variation between the values calculated for each object.
Small values mean more similar trajectory distances and less variance across them.

\subsubsection{Table Setup Task}
For the Table Setup Task measurements like each objects' interaction time as well as the overall object interaction time, the task completion time, idle time and grasp amount were analyzed.
The graspable objects in this scene were a knife, a spoon, a  fork, a plate, and an orange juice carton. 
The focus of this task is to measure the control and outcome differences for a series of pick and place tasks.
The results are summarized in Table \ref{tab:ts_interaction_times}.
Although not every object showed a statistically significant difference in grasp duration, the overall trend indicates that participants using the Manus motion capture gloves had the longest interaction and grasp times among all conditions.
For task completion time, participants using controllers were statistically significantly faster than those using the motion capture gloves.
They also had the lowest idle time, although no statistically significance could be found here ($\text{M-C}: p = 0.139, \text{H-C}:p=0.053$).
Considering the measured trajectory distances, regarding DTW in the Median and the standard deviation, C has significant lower values than the other controls. 
M has higher trajectory variety than both other controls and H has a higher variety than C.
Therefore, the typical median distance to the other objects within the M control is significantly higher than the values for the same object but using another control.

\begin{table*}[h]
    \centering
    \begin{adjustbox}{max width=\textwidth, center}
    \begin{tabular}{lcccllllc}
    
        \toprule
        \textbf{General Metrics} & \textbf{M} & \textbf{H} & \textbf{C} & \textit{p}-value & MC (\textit{p}, $\delta$) & MH (\textit{p}, $\delta$) & HC (\textit{p}, $\delta$) & Result \\
        \midrule
        Balance & 0.63 & 0.42 & 0.41 & .048$^*$ & .056,\hphantom{0} 0.49 & .141, \hphantom{0}0.33 & .488, \hphantom{0}0.12 & M $>$ C \\
        Complete Task Time & 124.17 & 122.31 & 95.16 & .006$^*$ & .009$^*$, 0.59 & .923, \hphantom{0}-0.02 & .009$^*$, 0.49 & H, M $>$ C \\
        Idle Time & 89.49 & 95.84 & 71.89 & .039$^*$ & .139,\hphantom{0} 0.35 & .336,\hphantom{0} -0.18 & .053,\hphantom{0} 0.46 & – \\
        Grasp Count & 8 & 6 & 7 & .003$^*$ & .101, \hphantom{0}0.37 & .001$^*$,\hphantom{0}0.60 & .114, \hphantom{0}-0.31 & M $>$ H \\

        &&&&&&&&\\
        \toprule
        \textbf{Object Grasp Time (s)} & \textbf{M} & \textbf{H} & \textbf{C} & \textit{p}-value & MC (\textit{p}, $\delta$) & MH (\textit{p}, $\delta$) & HC (\textit{p}, $\delta$) & Result \\
        \midrule
        Knife & 9.4 & 4.5 & 4.8 & .024$^*$ & .027$^*$, 0.48 & .079, \hphantom{0}0.39 & .567, \hphantom{0}0.12 & M $>$ C \\
        Dessert Spoon & 6.5 & 4.1 & 4.2 & .142 & – & – & – & – \\
        Fork & 8.0 & 4.9 & 3.8 & .065 & – & – & – & – \\
        Plate & 9.6 & 4.2 & 3.5 & .016$^*$ & .016$^*$, 0.51 & .067, \hphantom{0}0.39 & .436, \hphantom{0}0.16 & M $>$ C \\
        Orange Juice & 13.9 & 7.1 & 7.1 & .042$^*$ & .084, \hphantom{0}0.41 & .084, \hphantom{0}0.42 & .927, \hphantom{0}0.03 & – \\
        Overall Grasp Time & 36.07 & 25.23 & 22.13 & \(< .001\)$^*$ & \(< .001\)$^*$, 0.75 & \(< .001\)$^*$, 0.65 & .330, \hphantom{0}0.20 & M $>$ C, H \\

        &&&&&&&&\\
        \toprule



        \textbf{DTW Median} & \textbf{M} & \textbf{H} & \textbf{C} & \textit{p}-value & MC (\textit{p}, $\delta$) & MH (\textit{p}, $\delta$) & HC (\textit{p}, $\delta$) & Result \\
        \midrule
        Knife & 54.13 & 41.3 & 18.2 & \(< .001\)$^*$ & \(< .001\)$^*$, 0.85 & \(< .001\)$^*$, 0.69 & \(< .001\)$^*$, 0.68 & M $>$ H $>$ C \\
        Dessert Spoon & 50.14 & 33.08 & 17.74 & \(< .001\)$^*$ & \(< .001\)$^*$, 0.91 & \(< .001\)$^*$, 0.84 & \(< .001\)$^*$, 0.67 & M $>$ H $>$ C \\
        Fork & 43.61 & 31.8 & 21.44 & \(< .001\)$^*$ & \(< .001\)$^*$, 0.81 & \(< .001\)$^*$, 0.75 & .003, 0.63 & M $>$ H $>$ C \\
        Plate & 65.31 & 38.92 & 17.56 & \(< .001\)$^*$ & \(< .001\)$^*$, 1.0 & \(< .001\)$^*$, 0.88 & \(< .001\)$^*$, 0.72 & M $>$ H $>$ C \\
        Orange Juice & 79.17 & 53.6 & 43.25 & \(< .001\)$^*$ & \(< .001\)$^*$, 0.98 & \(< .001\)$^*$, 0.59 & \(< .001\)$^*$, 0.58 & M $>$ H $>$ C \\

  &&&&&&&&\\
        \toprule

         \textbf{DTW SD} & \textbf{M} & \textbf{H} & \textbf{C} & \textit{p}-value & MC (\textit{p}, $\delta$) & MH (\textit{p}, $\delta$) & HC (\textit{p}, $\delta$) & Result \\
        \midrule
        Knife & 44.35 & 23.43 & 16.56 & \(< .001\)$^*$ & \(< .001\)$^*$, 0.98 & \(< .001\)$^*$, 0.92 & \(< .001\)$^*$, 0.61 & M $>$ H $>$ C \\
        Dessert Spoon & 27.63 & 11.83 & 14.24 & \(< .001\)$^*$ & \(< .001\)$^*$, 0.99 & \(< .001\)$^*$, 1.0 & \(< .001\)$^*$, -0.62 & M $>$ C $>$ H \\
        Fork & 21.68 & 9.51 & 13.93 & \(< .001\)$^*$ & \(< .001\)$^*$, 0.85 & \(< .001\)$^*$, 1.0 & \(< .001\)$^*$, -0.72 & M $>$ C $>$ H \\
        Plate & 23.11 & 14.59 & 10.96 & \(< .001\)$^*$ & \(< .001\)$^*$, 1.0 & \(< .001\)$^*$, 1.0 & \(< .001\)$^*$, 0.67 & M $>$ H $>$ C \\
        Orange Juice & 26.73 & 29.49 & 15.03 & \(< .001\)$^*$ & \(< .001\)$^*$, 0.99 & .100, -0.27 & \(< .001\)$^*$, 0.99 & M, H $>$ C \\



        &&&&&&&&\\
        \toprule

        \textbf{DFD Median} & \textbf{M} & \textbf{H} & \textbf{C} & \textit{p}-value & MC (\textit{p}, $\delta$) & MH (\textit{p}, $\delta$) & HC (\textit{p}, $\delta$) & Result \\
        \midrule
        Knife & 0.33 & 0.41 & 0.26 & .010$^*$ & .115, 0.40 & .162, -0.30 & .006$^*$, 0.54 & H $>$ C \\
        Dessert Spoon & 0.32 & 0.33 & 0.27 & .041$^*$ & .102, 0.35 & .568, -0.09 & .044$^*$, 0.49 & H $>$ C \\
        Fork & 0.29 & 0.37 & 0.32 & .006$^*$ & .149, -0.30 & .003$^*$, -0.58 & .149, 0.35 & H $>$ M \\
        Plate & 0.43 & 0.39 & 0.28 & .052 & – & – & – & – \\
        Orange Juice & 0.60 & 0.59 & 0.64 & .831 & – & – & – & – \\
        
        &&&&&&&&\\
        \toprule

       \textbf{DFD SD} & \textbf{M} & \textbf{H} & \textbf{C} & \textit{p}-value & MC (\textit{p}, $\delta$) & MH (\textit{p}, $\delta$) & HC (\textit{p}, $\delta$) & Result \\
        \midrule
        Knife & 1.77 & 0.30 & 0.28 & \(< .001\)$^*$ & \(< .001\)$^*$, 0.82 & \(< .001\)$^*$, 0.81 & .330, 0.22 & M $>$ C, H \\
        Dessert Spoon & 0.38 & 0.16 & 0.26 & \(< .001\)$^*$ & \(< .001\)$^*$, 0.76 & \(< .001\)$^*$, 0.91 & \(< .001\)$^*$, -0.85 & M $>$ C $>$ H \\
        Fork & 0.13 & 0.10 & 0.27 & \(< .001\)$^*$ & \(< .001\)$^*$, -0.94 & .001$^*$, 0.52 & \(< .001\)$^*$, -1.0 & C $>$ M, H \\
        Plate & 1.36 & 0.22 & 0.20 & \(< .001\)$^*$ & \(< .001\)$^*$, 1.0 & \(< .001\)$^*$, 1.0 & .026$^*$, 0.37 & M $>$ C, H \\
        Orange Juice & 0.33 & 0.38 & 0.21 & \(< .001\)$^*$ & .001$^*$, 0.70 & .102, -0.33 & \(< .001\)$^*$, 0.77 & M, H $>$ C \\

        \bottomrule
    \end{tabular}
    \end{adjustbox}
    \vspace{0.1cm}
    \caption{Median values for the evaluated metrics in the table setup scenario are reported for M (Manus hands), H (controller with hand visualization), and C (controller without hand visualization). In addition, the table presents p-values from the Kruskal–Wallis omnibus test, as well as results from pairwise comparisons with Holm correction combined with Cliff’s Delta effect sizes.}
    \label{tab:ts_interaction_times}
\end{table*}

\subsubsection{Dishwasher Task}
The Dishwasher Task, like the Table Setup Task, involved a sequence of pick-and-place interactions, but with an added focus on precision placement inside the dishwasher racks.
Table \ref{tab:dw_interaction_times} shows the numeric results of the analysis.
The Dessert Spoon had a significantly longer grasp duration when using Manus gloves compared to the other two conditions ($p < 0.01$). 
While the data shows that the overall grasp time for the Manus gloves were higher than the other controls consistent for nearly all objects, the difference did not reach statistical significance.
No significant differences were observed in task completion time, idle time, or grasp count across conditions, suggesting that all ICs allowed participants to complete the task at a similar pace.

When analyzing the trajectory distances, the Manus gloves consistently showed higher variability, indicating that participants had more varied motion paths when using them compared to the other ICs.
The Hand visualization also displayed increased variability compared to the Controller visualization, although to a lesser extent.
The Controller visualization exhibited more consistent trajectories with significantly lower DTW median and standard deviation values compared to both other controls.
And, although not always significant, the controller visualization had the lowest median distances regarding DFD median for all items besides the bread knife.

\subsubsection{Cutting Task}
The cutting task where the participants had to cut the bread was already considered in our prior work\cite{helmert2025semantic}.
In this task the focus was not set on the correct endposition of the objects like in the dishwasher or the table setup task.
Instead the movements during the grasping interactions were important.
Therefore we also set a focus on the semantics of the trajectory.
We analyzed if the order and amount of interactions were correct and if there are unnecessary or wrong interactions.
For this the moving trajectories of all objects were segmented as described in section \ref{subsec:trajectory_analysis}. 
Each segment has an own subtask, like grasping the knife, going to cut bread, cutting bread, or releasing the knife.
The subtask changes with every segment change.
The resulting optimal semantic trajectory for this task is then 
$$(\{\},\{Grasp\},\{Grasp,Cut\},\{Grasp\},\{Grasp,Cut\},$$
$$\{Grasp\},\{Grasp,Cut\},\{Grasp\},\{\})$$
which resembles first, by grasping the knife, followed by three cuts and then by placing the knife down.
We already found out that, by comparing the actual trajectories recorded, the trajectory's quality significantly differed between the control-visualization combinations \cite{helmert2025semantic}.
The concrete values for the following analysis are displayed in Table \ref{tab:ct_interaction_times}.
The Manus hand did significantly less unnecessary actions than the other controls and the Levenshtein distance was the lowest for this control.
Although not significant, the median cut count for the controller based controls was 4. 
Therefore at least half of all participant did too much cuts, although it was clearly stated that the bread should be cut exactly 3 times.
We also considered a spatial analysis of the trajectories, and used the Discrete Frechet (DFD)\cite{eiter1994computing} or Dynamic Time Warping (DTW) \cite{kassidas1998synchronization} as similarity measures here.
We applied these measures then for the cutting segment subtask and found out that the participants using the Manus motion capture gloves had significantly higher similarity between the trajectories than user of the other groups, indicating that the intuition of how this task is done, might be clearer to participants who used the the motion capture gloves than to ones using the other controls.
Again, like for the same knife in the dishwasher scenario, it had the lowest variation.

\subsubsection{Cleaning Task}
The cleaning task required participants to use a sponge to wipe a designated area on the table.
Unlike the table setup and dishwasher tasks, this task emphasized repetitive motions and required precise coverage of the surface rather than object placement. 
The key measurements for this task included grasp time, total task time, total cleaning time, idle time, number of grasps, and the number of distinct cleaning actions. The results are summarized in Table \ref{tab:cl_interaction_times}.
There were no significant differences in the overall grasp time or task time among the different controls.
The total time spent cleaning, which represents the sum of all cleaning actions, also showed no statistical significance across conditions ($p = 0.7001$).
Idle time, i.e the time no action was currently ongoing after starting the the first, was consistently zero across all conditions, indicating that participants did not release the sponge which is consistent to the grasp amount of one across all conditions.
The amount of cleaning actions shows a trend ($p = 0.07542$) regarding C having the lowest amount of cleaning actions.
A cleaning action started when the sponge entered a small the area that starts about $1cm$ above the surface and ends about $1cm$ below.
The cleaning action stopped when the sponge leaved that area.
While inside and the cleaning process is active, the sponge cleaned the dirty surface below.
When more than 90\% of the surface was cleaned, the tasks was considered successful, which also was automatically indicated to the users.
Therefore, a higher number of cleaning actions suggests that participants struggled to maintain the sponge within the designated area, leading to frequent re-engagements.
Since the task was to clean one surface, the optimal sequence of actions follows:
$$\{\},\{Grasp\},\{Grasp,Clean\},\{Grasp\},\{\}$$
where the cleaning is done releasing the sponge once and without ever leaving the cleaning area before the cleaning was finished.
While it did not count this as an error or unnecessary movement it can be seen as lack of accuracy, holding the sponge to far up or down.
Releasing the grasp, however, was considered an unnecessary movement or error.
This would indicate problems holding the sponge or inconsistencies in the task description or execution.
Overall, while the results did not show statistical differences, the cleaning action count suggests that using H control may have caused more difficulty in maintaining precise movements while the C control had the least difficulty doing so.
However, further analysis may be necessary to confirm this hypothesis.

\subsubsection{Pouring Task}
The pouring task required participants to pick up and tilt a pot and a milk pitcher to pour liquid in form of small balls into a designated container each.
Unlike the previous tasks, this task involved a continuous movement combined with having the correct orientation to prevent spilling.
Participants had the option to grasp the pot with one or both hands, but a two-handed grasp was not mandatory.
The key measurements for this task included overall grasp time, total task time, idle time, grasp count, and object-specific grasp counts.
The results are summarized in Table \ref{tab:p_interaction_times}.
There were no significant differences in overall grasp time ($p = 0.2375$) or total task time ($p = 0.3409$) among the different control conditions.
The idle time also showed no significant difference ($p = 0.3341$), and resemble here mostly the time users took switching from the first pouring to the second.
The grasp count varied slightly across conditions, but no significant differences were found.
We observed that users of all controls sometimes grasped the pot with both hands, the ones using Manus did it more often.
However, since this type of grasps was not possible in the current implementation, the pot was only attached to one of the hand and sometimes caused some confusion.
As a design implication would here be that having a natural control causes user to do more natural movements which in turn need to be designed and correctly implemented before.
Importantly, we observed no user-caused errors during the pouring task, and the recorded movement trajectories did only significantly differ regarding the DTW distance but not the general metricts, indicating that this task was equally to solve across all tested ICs.

\subsubsection{Pointing and Rating Task}
\label{subsec:results_pointing}
In the pointing scene, the participants had to point at the center of 6 targets at different directions and distances.
We recorded the precision of the pointing interaction by measuring the distance between the point they pointed at and the actual center of the object.
Precise pointing is crucial for interactions that rely heavily on accuracy, such as targeted selection or precise gesture-based communication. 
Calculating the precision across different sessions in the future also helps tracking progress of fine motor based coordination.
With
$$D_{k,i,j} = |(t_{j})-(p_{i,j,k})|, i \in \{1,...,n,\},j \in \{1,..,6\}$$ 
being the Euclidean distance between the center of the $j$-th target $t_j$ and the position $p_{k,i,j}$ participant $i$ with control condition $k$ pointed at, we can calculate the mean distances for any fixed control $K$ and participant $I$ with
\[\Bar{D}_{K,I} = \frac{1}{6}\sum_{j=1}^6{|t_{I,j,K}-p_{I,j,K}|}\]
For only a fixed control $K$ the mean distances and standard deviations are defined as
\[\Bar{D}_K = \frac{1}{120}\sum_{i=1}^{20}\sum_{j=1}^6{|t_{i,j,K}-p_{i,j,K}|}\]
and
\[\sigma_K = \sqrt{\frac{1}{120}\sum_{i=1}^{20}\sum_{j=1}^6 ( |(t_{i,j,K})-(p_{i,j,K})|- \Bar{x}_K)^2} \]
($\Bar{D}_M = 9.0, \sigma_M =3$, $\Bar{D}_C = 6.8, \sigma_C = 2.3$,$\Bar{D}_H = 6.1, \sigma_H = 1.75$, all distances in $cm$).
For the distances per user we did Levene's test for homogeneity of variances.
It shows that the null hypothesis of homogeneity of variances holds ($p > 0.1$).
The Shapiro-Wilk normality test confirms that normality also holds ($p > 0.4$).
Therefore a One-way ANOVA test can be applied. It showed that there was a statistically significant difference in mean distances between at least two groups (F$(2, 57) = 8.15, p < 0.001$) and independent-samples t-tests can be applied.
The results indicated that participants using M performed worse than both other controls, the C ($p>0.01$) and the H ($p<0.001$) control.
This shows that in our experiments using a controller yields a higher accuracy when using a button to trigger the pointing action compared to the pointing gesture used by the motion capture hands.

\section{Discussion}
\label{sec:discussion}
This section discusses our findings in light of the research questions.
We first reflect on user experience (RQ1), then analyze task execution and interaction behavior (RQ2).

In this work, we created a virtual kitchen environment in which users executed realistic tasks.
We compared different interaction modalities and devices for these tasks in order to evaluate their performance regarding user experience and task performance.
Regarding user experience we found out that regarding the NASA TLX, the SUS and user comments, no significant differences were found and no control was really standing out. This also confirms the results of Khundam et al. \cite{DBLP:KhundamVPHN21} but also means that while having nothing in ones hand when using hand tracking is not as problematic as described by Johnson et al. \cite{DBLP:conf/hci/JohnsonFPEW23}.
The lack of significant differences suggests that, regarding user experience and workload in general, any of these controls can be selected for lay users (\textbf{RQ1}). \\
However, since the subjective experience alone does not distinguish between the input methods, it becomes more relevant to examine whether specific ICs lead to measurable differences in task execution or user behavior.
The results of our study also provide insights into how different ICs influence interaction performance for lay users in various kitchen-related tasks.
While not all tasks showed statistically significant differences, certain trends and observations suggest that the visualization and control method can impact user behavior and task efficiency.
Table~\ref{tab:summary_results} summarizes our key findings across tasks, visualizations, and control types with respect to \textbf{RQ2} which will be explained in more details in the following sections.

{
\newcolumntype{P}[1]{>{\raggedright\arraybackslash}p{#1}}

\renewcommand{\arraystretch}{1.2}
\rowcolors{2}{gray!10}{white} 
\begin{table*}[t]
\centering
\begin{adjustbox}{max width=\textwidth}
\begin{tabular}{P{1.8cm} P{2.3cm} P{4cm} P{5cm} P{2.4cm}}
\toprule
\textbf{Task}  & \textbf{Semantic Accuracy} & \textbf{Trajectory Similarity} & \textbf{Performance / Errors / Observations} & \textbf{Recommended IC} \\
\midrule

Table Setup  & H lower grasp count than M* and C & C lowest median DTW for all objects; C lowest median DFD for some objects;  M has a high variation & C faster than H\textsuperscript{*} and M\textsuperscript{*}; C has lowest idle time; M user used both hands more than C\textsuperscript{*} and H; & C or M depends \\

Dishwasher  & M slightly lower overall grasp counts & C lowest median DTW for most objects*; C lowest DTW SD for all objects*; M highest variation for most objects* &  No significant time/grasp count differences; M longer spoon grasp (smallest item)& C \\

Cutting  & M lowest Levenshtein distance\textsuperscript{*} &  M lowest median DTW/DFD\textsuperscript{*}; M lowest DTW/DFD SD\textsuperscript{*}; & H/C had too many cuts; H/C often regrasped the knife & M \\

Cleaning & C lowest Levenshtein distance &  C lowest DTW\textsuperscript{*}; C highest DFD\textsuperscript{*};& H had most cleaning actions & C \\

Pouring & H lowest grasp count &  Mixed values;  M mostly highest DTW/DFD\textsuperscript{*} and H mostly lowest\textsuperscript{*}  &  M tried many two-hand grasps on the pot  & All similar \\

Pointing \& Rating & Not applicable & Not applicable & M had the worst precision*; C and H often forgot how to point or rate & M or C/H depends \\

\bottomrule
\end{tabular}
\end{adjustbox}
\vspace{0.2cm}
\caption{Summary of performance, trajectory similarity, and errors across tasks and Input Configurations (ICs). C = Controller with controller visualization, H = Controller with hand visualization, M = Motion capture gloves. Entries with \textsuperscript{*} indicate statistically significant differences and depends means that it mostly depends on the objective e.g. natural user behavior, efficient solving or low variation.}
\label{tab:summary_results}
\end{table*}
}

\subsection{Segmentation Approach}
To address \textbf{RQ2} — how different ICs influence user behavior and task execution — we analyzed the recorded interaction data through a semantic trajectory segmentation approach.
Instead of treating a task as a single movement sequence, we split it into subtrajectories based on semantic transitions (e.g., reaching, grasping, cutting, releasing).
This allowed us to compare specific subtasks across the ICs and participants with more granularity than traditional path-based metrics.
Additionally, the semantic segmentation approach provided two main benefits.
First, it enabled a structured and readable overview of task execution, highlighting which subtasks were completed as intended.
Second, it allowed for subtask-specific similarity and error analysis (e.g., comparing only cutting segments rather than the full knife trajectory).
Although we found that segmentation enables more meaningful comparisons, it also has limitations:
Different participants might use different strategies (e.g., repositioning objects or sawing versus chopping bread), which makes absolute spatial alignment difficult.
Therefore, further refinement of similarity measures (e.g., invariance to initial offsets or non-task-relevant movement) remains an area for future work.
Nonetheless, our current approach provided a strong foundation for analyzing interaction quality across modalities.

\subsection{Impact of the Input Configuration on Task Performance}
To further address \textbf{RQ2}, we investigated how the different ICs influenced user behavior and performance across various task types.
We observed distinct patterns depending on whether the task was goal oriented (focused on outcomes) or manner oriented (focused on the way of execution).
\textbf{In goal-oriented tasks} such as the table setup or dishwasher scene, efficiency and precision in object placement were central. 
Here, the controller with controller visualization (C) consistently had faster task completions and showed the lowest idle times and grasp durations.
It also produced more consistent trajectories, indicated by significantly lower DTW and DFD distances.
This suggests that direct, button-based input supports a faster and efficient execution for short goal-oriented tasks.
The longer grasp times of the motion capture gloves (M) may indicate a trade-off between natural interaction and efficiency, where the gloves provide a more realistic grasping experience but at the cost of additional handling time. 
These findings align with the results from Argelaguet et al. \cite{DBLP:conf/vr/ArgelaguetHTL16} and Kangas et al. \cite{DBLP:journals/mti/KangasKMJR22}.

\textbf{In manner-oriented tasks} such as cutting, pouring, and cleaning, the focus was on continuous motion and interaction quality rather than completion time.
In these scenarios, the motion capture gloves (M) resulted in more natural execution patterns and more semantically accurate trajectories (e.g., fewer unnecessary grasps and better alignment with the intended action sequence).
In the cutting task, as an example, participants using gloves performed significantly closer to the optimal semantic and spatial trajectory, which is reflected by the lowest Levenshtein distances and highest spatial trajectory similarity across this group.
This implies that more natural interactions here lead to more efficient task executions for these tasks.
However, this advantage was not consistently observed in all manner-oriented tasks which means further research might be necessary to confirm or reject this hypothesis.
For cleaning, controller users (C) showed slightly more precise movements, although the differences were not statistically significant
For pouring it could be the fact that it can be done faster with less room to make errors. 

\textbf{In gesture-focused tasks}, such as pointing and rating, motion capture gloves were perceived as more intuitive and helped participants memorize gestures more easily compared to controller-based configurations.
However, the precision of gesture execution with M was significantly lower.
This is likely due to the continuous nature of gesture recognition and that its thresholds are not as transparent to the users as for buttons on controller which in turn makes it harder for users to know when a gesture was successfully.

Taken together, these findings show that task characteristics strongly interact with the IC:
Controllers support efficient and consistent behavior in goal-oriented tasks, gloves enable more natural and semantically correct execution in manner-oriented tasks, and gesture-based interactions benefit from the intuitiveness of gloves but at the cost of lower pointing precision.

\subsection{Precision and Errors in Different Tasks}
A further perspective on \textbf{RQ2} considers semantic execution accuracy and interaction precision.
Here, we identified several patterns related to the IC-specific quality of task execution.

\noindent \textbf{Semantic accuracy} was most clearly reflected in the cutting task.
The motion capture gloves (M) led to significantly lower Levenshtein distances compared to the controller-based configurations (Median Levenshtein distance = 0 for M and versus 3 for H and C), indicating a closer match to the optimal semantic executions (i.e., optimal sequence and number of grasps and cuts).
Participants using controllers often performed too many grasps or cuts, leading to suboptimal action sequences.
This finding highlights how naturalistic ICs promote more and manner-aligned interaction behavior.
These results indicate that tasks requiring fine motor stability over a longer time are affected by the IC and natural interactions solely relying on gestures are advantageous in this case.

\noindent \textbf{Interaction precision} was assessed in the pointing and rating task.
Here, participants using M showed significantly worse pointing accuracy than those using C or H.
This is likely due to the continuous nature of gesture recognition and the lack of a clear trigger event when using gesture input.
In contrast, pressing a button on a controller provides discrete and understandable interaction timing, resulting in higher spatial precision.

\noindent \textbf{Errors and misinteractions} were observed across all ICs and task, but their nature differed.
Participants using the H configuration (controller with hand visualization) sometimes struggled to grasp small objects or accidentally released items, especially during longer interactions.
Since grasping in this configuration is performed by squeezing the controller, we assume that the applied force weakens over time, occasionally leading to unintentional releases.
In contrast, participants using M (motion capture gloves) occasionally triggered unintended grasps, as the grasp gesture can be performed too easily and may be activated unintentionally.
These findings suggest that each IC has its own interaction sensitivities, which may influence training or learning outcomes depending on the target application.
The pouring task showed almost no user-caused errors across all configurations—despite many M participants attempting unsupported two-handed grasps on the pot.
This indicates that pouring does not require strong visual or haptic feedback for successful execution and can be performed equally well with all configurations.
The cleaning task, however, showed a trend in the number of cleaning actions ($p = 0.07542$), suggesting that users using the H configuration struggled to maintain precise vertical control.

\noindent In summary, ICs differ not only in performance metrics but also in the type and frequency of errors.
Naturalistic input enable more semantically correct executions, while controller-based input offers often but not always higher interaction precision and less variability in task executions.

\subsection{Grasping and Task Adaptation}
Another relevant aspect related to \textbf{RQ2} is how participants adapted their grasping strategies depending on the IC.
Participants using motion capture gloves (M) frequently used both hands in natural ways — for example, opening a drawer with one hand while retrieving an object with the other.
In the pouring task, many attempted to grasp the pot with both hands, even though only single-handed grasps were supported.
This suggests that naturalistic input promotes more intuitive interaction behavior.
By contrast, participants using controllers (H and C) often tended to rely on single-handed actions, likely due to the button-based input and missing naturalness of interactions.
These observations indicate that the ICs not only influence performance but also users’ interaction strategies.

\subsection{Implications for VR-based Applications}
Our findings offer practical implications for selecting ICs in VR-based systems, depending on the intended application goals as shown in Figure~\ref{tab:summary_results}.
If efficiency and consistency are critical — for example, in training scenarios focused on speed and accuracy — controller-based input (C or H) is advantageous due to faster execution and lower trajectory variability.
In contrast, if the goal is to support more intuitive and embodied interaction behavior — such as in rehabilitation or exploratory learning environments — motion capture gloves (M) offer benefits through more natural execution and increased use of both hands.
However, the higher variability and presence of unnecessary actions with naturalistic input can be a disadvantage in scenarios where clean and structured demonstrations are required.

\section{Limitations and caveats}
\label{sec:limits}
Although we found many significances it is not always clear where they originate from.
The median distances are different, but even if they have these low p-values and good effect sizes, it is not clear if these variations really make a difference in the quality.
Also, some of these like the distance differences might be the consequence of grasping or other problems. 
Additionally, the distance metrics are both not complete metric for overall similarity.
Each of them has different strengths and weaknesses and in overall none of these distance measures gives a complete and representative metric for the similarity of the trajectory. 
There are too many different factors, that influence the outcome. 
Therefore, we used them complementary to other measurements. 

While DTW is good in finding similarity in overall shape and timings even if they are warped or shifted, the DFD is more strict. 
It calculates the minimal bottleneck and is susceptible to outlier or sections where trajectories, even shortly, vastly differ.
It penalizes spatial differences harder, independent of the shape similarity of the rest of the trajectory.

While our study provides valuable insights, several limitations should be considered.
First, the lack of significant differences in certain tasks may be due to sample size limitations or task design constraints.
Additionally, an important direction for future research is investigating long-term adaptation.
Users might initially struggle with motion capture gloves due to their different interaction paradigm, but they could potentially become more efficient with continued use.
The same might count for the other controls in aspect they were worse.
A longitudinal study could help determine whether these initial differences persist or diminish over time.

\section{Conclusion}
\label{sec:conclusion}

In this work, we explored how different input configurations(ICs)---defined by both hardware and their virtual visualization---affect user experience and task execution in virtual reality.

\textbf{RQ1} asked which configuration provides the best user experience in terms of usability and workload.
Our results show that all three configurations---motion capture gloves, controllers with hand visualization, and controllers with controller visualization---achieved similarly high usability and low to medium workload ratings.
This suggests that all input configurations are suitable for lay users in terms of subjective experience.

\textbf{RQ2} focused on how input configurations influence user behavior and interaction quality.
Here, we found clear differences depending on the task type.
Controller-based configurations supported faster and more consistent performance in goal-oriented tasks, while motion capture gloves led to more semantically accurate, natural and more variated movement behaviors in manner-oriented tasks like cutting or cleaning.

As a secondary contribution, we proposed a semantic trajectory framework that segments tasks into subtasks and enables fine-grained comparisons of execution quality.
This approach enables the analysis of execution structure, error patterns, and interaction consistency across input configurations, and in the future may create adaptive feedback systems that analyze movements in real time.

Taken together, these findings highlight the trade-offs between efficiency and naturalism in VR interaction design.
Depending on whether the goal is precise execution, realistic behavior, or intuitive interaction, different input configurations may be more appropriate.
Future work should further explore how these trade-offs behave in long-term use and real-world applications.

\backmatter

\section*{Declarations}

\subsection*{Funding}
Jonas Dech and Abhijit Vyas acknowledge support by the Deutsche Forschungsgemeinschaft (DFG, German Research Foundation) – Project-ID 329551904 – SFB 1320 "EASE – Everyday Activity Science and Engineering".
\subsection*{Ethics approval and consent to participate} 
The study was approved by the Ethics Committee of the Bielefeld University (approval no. 2024-027). All participants provided written informed consent for participation in the study and for the collection, processing, and storage of the collected data.

\bibliography{sn-bibliography}

\newpage

\begin{appendices}

\section{Data Sheets}\label{secA1}

This section contains the data sheets for the statistical tests made for the different scenarios.

{
\renewcommand{\arraystretch}{1}
\begin{table*}[h]
    \centering
    \begin{adjustbox}{max width=\textwidth, center}
    \begin{tabular}{lcccllllc}

        \textbf{General Metrics} & \textbf{M} & \textbf{H} & \textbf{C} & \textit{p}-value & MC (\textit{p}, $\delta$) & MH (\textit{p}, $\delta$) & HC (\textit{p}, $\delta$) & Result \\
        \midrule
        Balance & 0.75 & 0.53 & 0.77 & .123 & – & – & – & – \\
        Complete Task Time & 134.55 & 144.87 & 135.28 & .979 & – & – & – & – \\
        Idle Time & 94.70 & 112.05 & 104.95 & .786 & – & – & – & – \\
        Grasp Count & 11 & 13 & 14 & .484 & – & – & – & – \\
        
           &&&&&&&&\\

        \toprule
        \textbf{Object Grasp Time (s)} & \textbf{M} & \textbf{H} & \textbf{C} & \textit{p}-value & MC (\textit{p}, $\delta$) & MH (\textit{p}, $\delta$) & HC (\textit{p}, $\delta$) & Result \\
        \midrule
        Dessert Spoon & 14.88 & 7.39 & 7.28 & .003$^*$ & .004$^*$, 0.63 & .004$^*$, 0.60 & .998, -0.01 & M $>$ H, C \\
        Plate & 10.63 & 9.15 & 9.01 & .577 & – & – & – & – \\
        Big Bowl & 14.36 & 9.02 & 9.06 & .056 & – & – & – & – \\
        Bread Knife & 7.84 & 7.87 & 7.01 & .521 & – & – & – & – \\
        Overall Grasp Time & 40.90 & 33.54 & 35.05 & .142 & – & – & – & – \\
         &&&&&&&&\\
        \toprule

        \textbf{DTW Median} & \textbf{M} & \textbf{H} & \textbf{C} & \textit{p}-value & MC (\textit{p}, $\delta$) & MH (\textit{p}, $\delta$) & HC (\textit{p}, $\delta$) & Result \\
        \midrule
        Dessert Spoon & 213.6 & 199.47 & 131.93 & \(< .001\)$^*$ & \(< .001\)$^*$, 0.74 & .191, 0.28 & .007$^*$, 0.58 & M $>$ H $>$ C \\
        Plate & 192.19 & 207.11 & 164.61 & \(< .001\)$^*$ & \(< .001\)$^*$, 0.65 & .348, -0.16 & \(< .001\)$^*$, 0.80 & H $>$ M, C \\
        Big Bowl & 180.1 & 180.14 & 126.85 & \(< .001\)$^*$ & \(< .001\)$^*$, 0.96 & .200, 0.23 & \(< .001\)$^*$, 0.91 & M, H $>$ C \\
        Bread Knife & 146.55 & 181.37 & 146.65 & .022$^*$ & .388, -0.15 & .017$^*$, -0.58 & .092, 0.38 & H $>$ M \\

        &&&&&&&&\\
        \toprule
        \textbf{DTW SD} & \textbf{M} & \textbf{H} & \textbf{C} & \textit{p}-value & MC (\textit{p}, $\delta$) & MH (\textit{p}, $\delta$) & HC (\textit{p}, $\delta$) & Result \\
        \midrule
        Dessert Spoon & 72.44 & 69.87 & 54.93 & .004$^*$ & .004$^*$, 0.65 & .633, 0.06 & .013$^*$, 0.50 & M, H $>$ C \\
        Plate & 107.18 & 60.54 & 39.93 & \(< .001\)$^*$ & \(< .001\)$^*$, 1.0 & \(< .001\)$^*$, 0.94 & .006$^*$, 0.48 & M $>$ H $>$ C \\
        Big Bowl & 208.96 & 48.84 & 36.46 & \(< .001\)$^*$ & \(< .001\)$^*$, 1.0 & \(< .001\)$^*$, 1.0 & \(< .001\)$^*$, 0.64 & M $>$ H $>$ C \\
        Bread Knife & 129.91 & 59.01 & 54.3 & \(< .001\)$^*$ & \(< .001\)$^*$, 0.95 & \(< .001\)$^*$, 0.93 & .067, 0.36 & M $>$ H, C \\

        &&&&&&&&\\
        \toprule

        \textbf{DFD Median} & \textbf{M} & \textbf{H} & \textbf{C} & \textit{p}-value & MC (\textit{p}, $\delta$) & MH (\textit{p}, $\delta$) & HC (\textit{p}, $\delta$) & Result \\
        \midrule
        Dessert Spoon & 0.80 & 1.15 & 0.49 & \(< .001\)$^*$ & .020$^*$, 0.51 & .020$^*$, -0.53 & \(< .001\)$^*$, 0.80 & H $>$ M $>$ C \\
        Plate & 0.55 & 0.52 & 0.50 & .353 & – & – & – & – \\
        Big Bowl & 0.43 & 0.56 & 0.34 & \(< .001\)$^*$ & \(< .001\)$^*$, 0.79 & .190, -0.30 & \(< .001\)$^*$, 0.84 & M, H $>$ C \\
        Bread Knife & 0.40 & 0.52 & 0.73 & .003$^*$ & .002$^*$, -0.61 & .036$^*$, -0.51 & .246, -0.28 & C $>$ M, H \\
         
        &&&&&&&&\\
        \toprule
        \textbf{DFD SD} & \textbf{M} & \textbf{H} & \textbf{C} & \textit{p}-value & MC (\textit{p}, $\delta$) & MH (\textit{p}, $\delta$) & HC (\textit{p}, $\delta$) & Result \\
        \midrule
        Dessert Spoon & 0.28 & 0.39 & 0.32 & .124 & – & – & – & – \\
        Plate & 0.19 & 0.20 & 0.22 & .109 & – & – & – & – \\
        Big Bowl & 0.62 & 0.19 & 0.16 & \(< .001\)$^*$ & \(< .001\)$^*$, 0.76 & .002$^*$, 0.73 & .024$^*$, 0.54 & M $>$ H $>$ C \\
        Bread Knife & 0.27 & 0.26 & 0.28 & .616 & – & – & – & – \\

        \bottomrule

    \end{tabular}
    \end{adjustbox}
    \vspace{0.1cm}
    \caption{Median values for the evaluated metrics in the dishwasher scenario are reported for M (Manus hands), H (controller with hand visualization), and C (controller without hand visualization). In addition, the table presents p-values from the Kruskal–Wallis omnibus test, as well as results from pairwise comparisons with Holm correction combined with Cliff’s Delta effect sizes.}
    \label{tab:dw_interaction_times}
\end{table*}

}

{
\renewcommand{\arraystretch}{1}
\begin{table*}[h]
    \centering
    \begin{adjustbox}{max width=\textwidth, center}
    \begin{tabular}{lcccllllc}
        \toprule
        
        \textbf{General Metrics} & \textbf{M} & \textbf{H} & \textbf{C} & \textit{p}-value & MC (\textit{p}, $\delta$) & MH (\textit{p}, $\delta$) & HC (\textit{p}, $\delta$) & Result \\
        \midrule
        Task Time & 16.61 & 16.99 & 21.02 & .805 & – & – & – & – \\
        Grasp Time & 15.79 & 14.62 & 18.7 & .673 & – & – & – & – \\
        Idle Time & 0 & 4.6 & 1 & .084 & – & – & – & – \\
        Grasp Count & 1 & 2 & 2 & .127 & – & – & – & – \\
        Cut Count & 3 & 4 & 4 & .171 & – & – & – & – \\
        Levenshtein D. & 0 & 3 & 3 & .001$^*$ & .002$^*$, -0.57 & .001$^*$, -0.61 & .752, 0.06 & C, H $>$ M \\

        \toprule

        \textbf{DTW Median}&&&&&&&&\\
        \midrule
        Bread Knife & 16.24 & 38.38 & 36.34 & \(< .001\)$^*$ & \(< .001\)$^*$, -0.94 & \(< .001\)$^*$, -0.85 & .580, 0.15 & H, C $>$ M \\

        \toprule

        \textbf{DTW SD} &&&&&&&&\\
        \midrule
        Bread Knife & 7.02 & 17.2 & 20.41 & \(< .001\)$^*$ & \(< .001\)$^*$, -1.0 & \(< .001\)$^*$, -0.58 & \(< .001\)$^*$, -0.57 & C $>$ H $>$ M \\

        \toprule

        \textbf{DFD Median}&&&&&&&&\\
        \midrule
        Bread Knife & 0.15 & 0.32 & 0.26 & \(< .001\)$^*$ & \(< .001\)$^*$, -0.80 & \(< .001\)$^*$, -0.79 & .210, 0.27 & H, C $>$ M \\

        \toprule

        \textbf{DFD SD}&&&&&&&&\\
        \midrule
        Bread Knife & 0.08 & 0.33 & 0.33 & \(< .001\)$^*$ & \(< .001\)$^*$, -1.0 & \(< .001\)$^*$, -0.90 & .490, 0.17 & C, H $>$ M \\

        \bottomrule
    \end{tabular}
    \end{adjustbox}
    \vspace{0.1cm}
    \caption{Median values for the evaluated metrics in the cutting scenario are reported for M (Manus hands), H (controller with hand visualization), and C (controller without hand visualization). In addition, the table presents p-values from the Kruskal–Wallis omnibus test, as well as results from pairwise comparisons with Holm correction combined with Cliff’s Delta effect sizes.}
    \label{tab:ct_interaction_times}
\end{table*}
}

{
\renewcommand{\arraystretch}{1}

\begin{table*}[h]
    \centering
    \begin{adjustbox}{max width=\textwidth, center}
    \begin{tabular}{lcccllllc}
        \toprule
        \textbf{General Metrics} & \textbf{M} & \textbf{H} & \textbf{C} & \textit{p}-value & MC (\textit{p}, $\delta$) & MH (\textit{p}, $\delta$) & HC (\textit{p}, $\delta$) & Result \\
        \midrule
        Task Time & 16.58 & 15.51 & 18.67 & .346 & – & – & – & – \\
        Clean Time Sum & 3.93 & 3.36 & 3.11 & .700 & – & – & – & – \\
        Idle Time & 0 & 0 & 0 & .257 & – & – & – & – \\
        Grasp Count & 1 & 1 & 1 & .138 & – & – & – & – \\
        Clean Count & 6 & 8.5 & 4 & .075 & – & – & – & – \\
        Levenshtein Dist. & 12 & 14 & 6 & .143 & – & – & – & – \\
        \toprule

        Grasp Times&&&&&&&&\\
        \midrule
          \textbf{Sponge} &    16.04 & 14.93 & 17.05 & .447 & – & – & – & – \\
        \toprule
  
        DTW Median&&&&&&&&\\
        \midrule
        \textbf{Sponge} &26.08 & 26.3 & 20.73 & \(< .001\)$^*$ & \(< .001\)$^*$, 0.84 & .830, -0.00 & \(< .001\)$^*$, 0.75 & M, H $>$ C \\
        \toprule

        DTWsd &&&&&&&&\\
        \midrule
        \textbf{Sponge} &7.27 & 7.55 & 5.82 & .002$^*$ & .005$^*$, 0.56 & .682, -0.09 & .002$^*$, 0.58 & M, H $>$ C \\
        \toprule

        DFD Median&&&&&&&&\\
        \midrule
        \textbf{Sponge}& 0.39 & 0.39 & 0.42 & .027$^*$ & .043$^*$, -0.45 & .910, 0.04 & .043$^*$, -0.43 & C $>$ M, H \\

        \toprule

        DFD sd&&&&&&&&\\
        \midrule
        \textbf{Sponge}& 0.09 & 0.09 & 0.16 & \(< .001\)$^*$ & \(< .001\)$^*$, -0.89 & .750, -0.06 & \(< .001\)$^*$, -0.89 & C $>$ M, H \\

        \bottomrule
    \end{tabular}
    \end{adjustbox}
    \vspace{0.1cm}
    \caption{Median values for the evaluated metrics in the cleaning scenario are reported for M (Manus hands), H (controller with hand visualization), and C (controller without hand visualization). In addition, the table presents p-values from the Kruskal–Wallis omnibus test, as well as results from pairwise comparisons with Holm correction combined with Cliff’s Delta effect sizes.}
    \label{tab:cl_interaction_times}
\end{table*}
}

{
\renewcommand{\arraystretch}{1}

\begin{table*}[h]
    \centering
    \begin{adjustbox}{max width=\textwidth, center}
    \begin{tabular}{lcccllllc}
       
        \toprule
        \textbf{General Metrics} & \textbf{M} & \textbf{H} & \textbf{C} & \textit{p}-value & MC (\textit{p}, $\delta$) & MH (\textit{p}, $\delta$) & HC (\textit{p}, $\delta$) & Result \\
        \midrule
        Task Time & 21.6 & 21.64 & 24.94 & .341 & – & – & – & – \\
        Idle Time & 4.46 & 5.92 & 8.25 & .334 & – & – & – & – \\
        Grasp Count & 5 & 3 & 4.5 & .278 & – & – & – & – \\
        
        &&&&&&&&\\
        \toprule

        \textbf{Grasp Times (s)} & \textbf{M} & \textbf{H} & \textbf{C} & \textit{p}-value & MC (\textit{p}) & MH (\textit{p}) & HC (\textit{p}) & Result \\
        \midrule
        Pot & 9.62 & 7.81 & 9.38 & .251 & – & – & – & – \\
        Milk Pitcher & 9.55 & 7.01 & 9.73 & .086 & – & – & – & – \\
        Overall Grasp Time & 17.15 & 15.07 & 19.33 & .238 & – & – & – & – \\

        &&&&&&&&\\
        \toprule

        \textbf{DTW Median} & \textbf{M} & \textbf{H} & \textbf{C} & \textit{p}-value & MC (\textit{p}, $\delta$) & MH (\textit{p}, $\delta$) & HC (\textit{p}, $\delta$) & Result \\
        \midrule
        Pot & 29.26 & 15.8 & 20.81 & \(< .001\)$^*$ & .010$^*$, 0.51 & \(< .001\)$^*$, 0.80 & .009$^*$, -0.56 & M $>$ C $>$ H \\
        Milk Pitcher & 12.39 & 8.06 & 9.91 & \(< .001\)$^*$ & .015$^*$, 0.52 & \(< .001\)$^*$, 0.84 & .006$^*$, -0.59 & M $>$ C $>$ H \\

        &&&&&&&&\\
        \toprule

        \textbf{DTW SD} & \textbf{M} & \textbf{H} & \textbf{C} & \textit{p}-value & MC (\textit{p}, $\delta$) & MH (\textit{p}, $\delta$) & HC (\textit{p}, $\delta$) & Result \\
        \midrule
        Pot & 18.66 & 18.57 & 38.97 & \(< .001\)$^*$ & \(< .001\)$^*$, -0.79 & .720, 0.09 & \(< .001\)$^*$, -0.79 & C $>$ M, H \\
        Milk Pitcher & 23.21 & 18.23 & 11.19 & \(< .001\)$^*$ & \(< .001\)$^*$, 0.86 & .003$^*$, 0.71 & \(< .001\)$^*$, 0.87 & M $>$ H $>$ C \\

        &&&&&&&&\\
        \toprule
        \textbf{DFD Median} & \textbf{M} & \textbf{H} & \textbf{C} & \textit{p}-value & MC (\textit{p}, $\delta$) & MH (\textit{p}, $\delta$) & HC (\textit{p}, $\delta$) & Result \\
        \midrule
        Pot & 0.19 & 0.18 & 0.17 & .203 & – & – & – & – \\
        Milk Pitcher & 0.12 & 0.12 & 0.10 & .109 & – & – & – & – \\
        
        &&&&&&&&\\
        \toprule
        \textbf{DFD SD} & \textbf{M} & \textbf{H} & \textbf{C} & \textit{p}-value & MC (\textit{p}, $\delta$) & MH (\textit{p}, $\delta$) & HC (\textit{p}, $\delta$) & Result \\
        \midrule
        Pot & 0.21 & 0.16 & 0.40 & \(< .001\)$^*$ & \(< .001\)$^*$, -0.80 & \(< .001\)$^*$, 0.85 & \(< .001\)$^*$, -0.90 & C $>$ M $>$ H \\
        Milk Pitcher & 0.13 & 0.12 & 0.17 & \(< .001\)$^*$ & \(< .001\)$^*$, -0.88 & .013$^*$, 0.58 & \(< .001\)$^*$, -0.89 & C $>$ M $>$ H \\
               
        \bottomrule
    \end{tabular}
    \end{adjustbox}
    \vspace{0.1cm}
    \caption{Median values for the evaluated metrics in the pouring scenario are reported for M (Manus hands), H (controller with hand visualization), and C (controller without hand visualization). In addition, the table presents p-values from the Kruskal–Wallis omnibus test, as well as results from pairwise comparisons with Holm correction combined with Cliff’s Delta effect sizes.}
    \label{tab:p_interaction_times}
\end{table*}
}




\end{appendices}


\end{document}